\documentclass[twocolumn,showpacs,prc,aps,floatfix,reprint]{revtex4-1}

\usepackage{graphicx}
\usepackage{dcolumn}
\usepackage{bm}

\usepackage{hyperref}

\usepackage{amssymb}
\usepackage{braket}
\usepackage{graphicx}
\usepackage{dcolumn}
\usepackage{amsmath}
\usepackage[capitalize]{cleveref}
\usepackage{amsfonts}
\usepackage{color}
\usepackage{float}
\usepackage{booktabs}
\usepackage{diagbox}
\usepackage{multirow}
\usepackage{siunitx} 

\usepackage{placeins}
\usepackage{bm}
\usepackage{capt-of}
\begin{document}

\preprint{APS/123-QED}

\title{Extension of the dispersive optical model to improve the description of high-momentum components
}

\author{R. A. Ramon$^{1}$, M. C. Atkinson$^{2}$, and W. H. Dickhoff$^1$}

\affiliation{${}^1$Department of Physics,
Washington University, St. Louis, Missouri 63130}
\affiliation{${}^2$Department of Physics and Astronomy,
Haverford College, Haverford, PA 19041}

\date{\today}

\begin{abstract}
An improved treatment of high-momentum components in nuclei is introduced in the framework of the dispersive optical model (DOM). The well-established feature that the peak of the spectral function appears at higher excitation energy in the $A$-1 system with increasing momentum has so far not been successfully accounted for in the DOM. To achieve this feature, it is necessary to abandon the factorization of energy dependence and geometry of the DOM self-energy.
The volume absorption below the Fermi energy has thus been represented by a decreasing radius at larger missing energy implying that a numerical treatment of the dispersion relations is necessitated. 
Earlier DOM results for ${}^{48}$Ca are also improved with this approach, demonstrating that a small neutron skin can still be accompanied by protons having a larger high-momentum tail than neutrons.
\end{abstract}

\maketitle


\section{Introduction}
\label{sec:intro}
Properties of high-momentum nucleons form an integral part of the study of nuclei~\cite{Hen:2017}.
Important properties have been identified that elucidate the dominance of proton-neutron short-range and tensor interactions in generating high-momentum nucleons in the nucleus~\cite{Subedi2008,Duer:2018,Schiavilla:2007}.
\textit{Ab initio} studies of infinite asymmetric matter confirm that minority protons become more correlated than majority neutrons with increasing nucleon asymmetry~\cite{Frick:2005,Rios:2009,Rios:2014} due to the role of the tensor force.
Such results are also observed in finite nuclei~\cite{Hen:2014momentum,Duer:2019direct}.
However, despite what has been observed, 
recent results demonstrate that the story is more complicated in that protons in ${}^{48}$Ca appear to be only mildly (around 10\%) more correlated than in ${}^{40}$Ca in the high-momentum region of 400 to 700~MeV/c whereas the corresponding ratio for ${}^{54}$Fe to ${}^{48}$Ca is more like 1.5~\cite{Nguyen2026}. To better understand these surprising results, we theoretically investigate high-momentum content in finite nuclei.

The existence of high-momentum nucleons in finite nuclei is one of the many observable effects of many-body correlations in nuclear physics. A consistent picture of these correlations is provided within a Green's function framework in which spectral functions provide the likelihood to find a nucleon with a particular momentum and energy within finite nuclei. Spectral functions can be determined not only from theoretical calculations~\cite{Dickhoff04}, but also from nucleon knockout experiments. Indeed, $(e,e'p)$ reactions reveal that valence proton orbits are only present at their presumed location for 60-70\%~\cite{Lapikas93}.
Compensation of this strength occurs both by admixing nearby single-particle strength associated with long-range (surface) correlations as well as high-momentum components generated by the role of tensor and short-range correlations.

The dispersive optical model (DOM) provides an empirical representation of the nucleon self-energy within the framework of Green's function theory~\cite{Mahaux:91,Dickhoff:2017,Dickhoff:2019,Exposed!} that can provide a complete picture of these correlations and address the presence of high-momentum nucleons, at least in principle. In the DOM framework, the imaginary part of the optical potential in the positive energy domain describes the damping of the quasi-particle states and the elastic scattering properties, while in the negative energy domain the imaginary potential describes the damping or spreading of the quasi-hole strengths, \textit{i.e.}, the fragmentation, as well as the appearance of high-momentum nucleons. The imaginary parts in both domains contribute to the real part through a subtracted dispersion relation.

According to the independent particle model (IPM), the spectroscopic factor corresponding to a quasi-hole peak is 1. However, the  $(e,e'p)$ reaction extracts for valence holes a spectroscopic factor of 0.6-0.7~\cite{Kramer:1989,Lapikas93}, \textit{i.e.}, 30-40\% of the strength is distributed outside the peak location. This redistribution of the strength of the quasi-hole peaks over a wide energy range is modeled by the imaginary part of the optical potential. 
In the negative-energy domain, it accounts for the damping of single-particle strength into more complicated many-body configurations below the Fermi energy, thereby contributing to its occupation in the ground state.
At positive energy, it generates the depletion of the strength of such valence orbits that require the admixture of states that are empty in the IPM.
The imaginary potential at large negative energies then partly represents the vehicle to absorb high-momentum strength to generate the correct particle number of the ground state~\cite{Exposed!}.

According to the original proposal of Mahaux and Sartor~\cite{Mahaux:91}, the
imaginary part of the self-energy in the DOM consists of an energy-dependent term multiplied by a potential representing its geometry.
This factorization facilitates the use of forms that allow analytical expressions~\cite{vanderkam2000analytical} for the dispersion integral to obtain the real part of the dynamic self-energy. 

Several improvements over the original form of the DOM~\cite{Mahaux:91} include a larger energy window~\cite{Charity06,Charity:2007}, nonlocality of the real Hartree-Fock-like potential~\cite{Dickhoff:2010,Mueller:2011}, and ultimately the full nonlocal representation of the imaginary part~\cite{Mahzoon:2014} to account for the charge density of the ground state.
This nonlocal version of the DOM was thus successful in forging a link between the domains of nuclear reactions and structure~\cite{Mahzoon:2014}. 
The resulting description of ${}^{40}$Ca
successfully described all the scattering observables, level structure, as well as the ground-state charge distribution, with the exception of the experimental spectral strength distribution at high missing energy and high missing momenta~\cite{Rohe:2004,Rohe:2004thesis}.

In Ref.~\cite{Mahzoon:2014}, the DOM spectral functions for $^{40}$Ca peak almost at the same missing energy $E_m$ for all missing momenta $p_m$, whereas the maximum of the experimental distorted spectral function $S(E_m, p_m)$ moves to the higher missing energies with increasing momenta~\cite{Rohe:2004}. 
From the perspective of the nucleon self-energy, a simple picture of the role of short-range correlations (SRC), dominated by np pairs~\cite{Hen:2014momentum,Duer:2018,Duer:2019direct}, 
arises when a high-momentum proton couples to an intermediate two-hole one-neutron particle state~\cite{Exposed!}.
Assuming that the two holes have almost no total momentum, the proton and neutron must have equal but opposite momentum.
The location of the corresponding removal strength of the proton particle then occurs at the average energy of the two holes minus the energy of the neutron particle.
The latter feature implies that when increasing the proton momentum, its peak location, determined by the vanishing of this energy denominator, must occur at more negative energy with a quadratic momentum dependence.
While the experimental maximum deviates from this simple picture, the location of the peak does exhibit a quadratic missing momentum dependence and therefore occurs at larger missing energy with increasing $p_m$~\cite{Rohe:2004thesis}.

We attribute the inability to capture this high-momentum behavior in the DOM spectral functions to the fact that 
the absorptive volume imaginary part of the self-energy fails to adequately describe the localized short-range correlation at larger binding energies, while the surface imaginary part, along with the Hartree-Fock (HF) part, describes the long-range correlation quite well near the Fermi energy. 
\textit{Ab initio} calculations of the self-energy for ${}^{16}$O~\cite{Muther:1994,Muether:1995} and ${}^{40}$Ca~\cite{Dussan:2011} exhibit a shrinking volume of the imaginary part with increasing missing energy in accord with the above simple argument.
Nuclear matter calculations of high-momentum spectral functions~\cite{RAMOS19891,BENHAR1989267} demonstrate this behavior very clearly, including the more recent self-consistent implementations of the Green's function method~\cite{Rios:2009,Rios:2014}.

Since many-body correlations evolve with energy, from long-range correlations near the Fermi energy to increasingly localized short-range correlations far away from the Fermi energy, it is physically a reasonable assumption that the volume-imaginary part of the self-energy should also evolve with energy.
This is particularly relevant for the volume absorption below the Fermi energy, which is the focus of the present paper.
\textit{Ab initio} calculations of the nucleon self-energy for ${}^{16}$O~\cite{Muther:1994,Muether:1995} and ${}^{40}$Ca~\cite{Dussan:2011} also exhibit this behavior and provide further motivation for the present study.

Another motivation for this study is the calculated momentum distribution of protons and neutrons in $^{48}$Ca~\cite{calleya2025investigating}. Although an earlier DOM fit~\cite{Mahzoon:2017} with a thicker neutron skin for $^{48}$Ca exhibited more high-momentum protons than neutrons, a later fit with a thinner skin generated more high-momentum neutrons than protons~\cite{calleya2025investigating}. However, knockout experiments~\cite{Duer:2018, Egiyan:2006Measurement} and \textit{ab initio} calculations for asymmetric matter~\cite{Rios:2009, Rios:2014} suggest an increased high-momentum content for the minority species in nuclei. Almost all high-momentum nucleons belong to the SRC pairs, and most of them are np pairs, known as ``np dominance''~\cite{Hen:2014momentum, Duer:2018, Duer:2019direct}, which is a direct consequence of the tensor force~\cite{Schiavilla:2007,Wiringa:2014}. Hence, in an asymmetric system with more neutrons than protons $(e.g.\ {}^{48}\mathrm{Ca}, ^{208}\mathrm{Pb} )$, protons should have more high-momentum components than neutrons.

To address these difficulties, we extend the DOM by introducing an energy-dependent geometry for the volume imaginary part of the self-energy below the Fermi energy that is documented in \textit{ab initio} calculations of the nucleon self-energy~\cite{Muther:1994,Muether:1995,Dussan:2011}. In this modification, we can no longer use the analytical expressions for the dispersive integral for the negative energy domain because the geometry and energy-dependent part cannot be separated anymore. Thus, this extension requires computationally expensive numerical calculations of the Cauchy principal-value (PV) integrals in the dispersion relation corresponding to the volume-imaginary term of the self-energy.
We emphasize that this extension is motivated by both experimental and theoretical considerations and is intended to improve the empirical description of high-momentum components in nuclei.

The paper is organized as follows. First, the theoretical framework behind the dispersive optical model in Sec.~\ref{sec:dom} is briefly reviewed. 
The implementation of energy dependence in the form factor of the volume-imaginary potential is discussed in Sec.~\ref{sec:Energy_dependence}. 
In Sec.\ref{sec:equivalence}, the numerical strategy is tested and shown to be equivalent to the analytic one in the appropriate limit.
Section~\ref{sec:bound_state_properties} shows the ability of the numerical extension to reproduce the bound-state properties of the standard DOM with analytical dispersion integrals. In Sec.~\ref{sec:volint}, the volume integrals of the imaginary part below the Fermi energy are compared between the numerical and standard version to demonstrate that the self-energy does not drastically change. 
It does however redistribute the location of the high-momentum strength to capture the SRC physics more realistically and thus generates better spectral functions, which is discussed in Sec.~\ref{sec:results}. 
Calculations performed using the 
standard DOM will be labeled as $\mathrm{DOM_A}$ and calculations completed by the numerical extension formulated in this article will be categorized as $\mathrm{DOM_N}$ throughout the text.
Some final conclusions are drawn in Sec.~\ref{sec:conclusion}. Details of the standard DOM parametrization are provided in App.~\ref{app:parametrization}. 
In App.~\ref{app:bound_state} bound-state results are presented, and in App.~\ref{app:scattering}. the relevant elastic scattering results.
Parameters are collected in App.~\ref{app:params}.

\section{Theory}
\label{sec:theory}

To clarify the extension of the dispersive optical model we first introduce the essentials of its standard form that includes a complete treatment of nonlocality~\cite{Mahzoon:2014} in Sec.~\ref{sec:dom}. 
Many of the expressions in this section are discussed in further detail in Refs.~\cite{Dickhoff:2017,atkinson2020dispersive,Exposed!}.
The proposal to include the energy dependence in the geometry of the volume imaginary part of the self-energy below the Fermi energy is discussed in Sec.~\ref{sec:Energy_dependence}. 

\subsection{Dispersive optical model}
\label{sec:dom}

In the many-body Green's function formalism, the irreducible self-energy, $\Sigma^*(\bm{r},\bm{r}';E)$, is a complex one-body potential which, in principle, is comprised of an infinite set of Feynman diagrams describing the propagation of an interacting nucleon through a nucleus based on a Hamiltonian containing relevant two- and three-body interactions~\cite{Exposed!}. In the DOM formalism, this complex one-body potential is parametrized as an optical potential which naturally extends to negative (bound) energies within the Green's function framework as first employed by Mahaux and Sartor~\cite{Mahaux91}. 
 The analytic structure of the nucleon self-energy allows one to utilize a dispersion relation, which relates the real part of the self-energy at a given energy to a dispersion integral of its imaginary part over all energies. The energy-independent correlated Hartree-Fock (HF) contribution~\cite{Exposed!} is removed by employing a subtracted dispersion relation with the Fermi energy used as the subtraction point~\cite{Mahaux91}
  \begin{align}
    \mathrm{Re}\ \Sigma^*(\bm{r},\bm{r}';E) &= \mathrm{Re}\
    \Sigma^*(\bm{r},\bm{r}';\varepsilon_F) \label{eq:dispersion} \\ -
    \mathcal{P}\int_{\varepsilon_F}^{\infty} \!\! \frac{dE'}{\pi}&\mathrm{Im}\
    \Sigma^*(\bm{r},\bm{r}';E')\left[\frac{1}{E-E'}-\frac{1}{\varepsilon_F-E'}\right] \nonumber
    \\ + \mathcal{P} \! \int_{-\infty}^{\varepsilon_F} \!\!
    \frac{dE'}{\pi}&\mathrm{Im}\
    \Sigma^*(\bm{r},\bm{r}';E')\left[\frac{1}{E-E'}-\frac{1}{\varepsilon_F-E'}\right],
    \nonumber      
 \end{align}
 where $\bm{r}$ also implies relevant discrete quantum numbers and $\varepsilon_F$ is the average Fermi energy which separates the particle and hole domains, 
 \begin{align*}
 \varepsilon_F = \frac{1}{2}\left(E_0^{A+1}-E_0^{A-1}\right),
 \end{align*}
 and $E_0^{A\pm1}$ represent the ground state energies of the $A\pm1$ nucleus~\cite{Exposed!}. 

Earlier work to include ground-state properties in the DOM demonstrated that it was not possible to reproduce the charge density without including a nonlocal imaginary part~\cite{Dickhoff:2010}.
We therefore continue to implement a nonlocal representation of the self-energy following
 Ref.~\cite{Mahzoon:2014} where $\Sigma_{\text{HF}}(\bm{r},\bm{r'}) = \mathrm{Re}\
    \Sigma^*(\bm{r},\bm{r}';\varepsilon_F) $ and the imaginary part 
 $\mathrm{Im}\ \Sigma(\bm{r},\bm{r'};E)$ are parametrized, and Eq.~\eqref{eq:dispersion} generates the energy dependence of the
 real part. The HF term consists of a volume term, spin-orbit term,  and a wine-bottle shape consistent with a microscopic analysis~\cite{Brida11}. The imaginary self-energy consists of volume, surface, and spin-orbit terms. 
 Nonlocality is represented using the Gaussian form as proposed in Ref.~\cite{Perey:1962}. 
 More details can be found in Ref.~\cite{atkinson2020dispersive}. The parameterization is presented in the App.~\ref{app:parametrization} 

To use the DOM self-energy for predictions, the parameters are fit through a weighted $\chi^2$ minimization of available elastic differential cross section data ($\frac{d\sigma}{d\Omega}$), analyzing power data ($A_\theta$),  reaction cross sections ($\sigma_r$), total cross sections ($\sigma_t$), charge density ($\rho_{\text{ch}}$), energy levels ($\varepsilon_{\ell j}$), particle number, the root-mean-square charge radius ($R_\mathrm{ch}$), and the energy of the ground state. The scattering calculations are performed in a partial wave basis (specifying orbital and total angular momentum with $\ell$ and $j$, respectively) using $\Sigma_{\ell j}^*(r,r';E)$ as an optical potential in the framework of $R$-matrix theory~\cite{Baye:2010}.
 All calculations are done in a Lagrange basis with 30 mesh points, where Legendre polynomials with a matching radius of 12 fm are used for scattering calculations and Laguerre polynomials are used for bound-state calculations~\cite{Baye_review,Baye:2010}.

We employ the Dyson equation to obtain the Green's function, $G_{\ell j}(r,r';E)$, from the DOM self-energy~\cite{Exposed!}.
 The particle number, binding energy, and charge density are all obtained from the hole spectral function which corresponds to the imaginary part of the Green's function, 
 \begin{equation}
    S^{(P,N)}_{\ell j}(r,r';E) = \frac{1}{\pi}\mathrm{Im}\ G^{(P,N)}_{\ell j}(r,r';E) ,
    \label{eq:spec}
 \end{equation}
where $P$ and $N$ denote proton and neutron, respectively.

The spectral function in momentum space for a given $\ell j$ can be obtained by employing the double Fourier-Bessel transformation of the imaginary part of the propagator expressed in coordinate space,
 \begin{eqnarray}
    \label{eq:momentum_space_S}
    S^{(P,N)}_{lj}(k;E)
    &=& \frac{2}{\pi^2}\int_{0}^{\infty} dr r^{2} \int_{0}^{\infty} dr' r'^{2} \\
    &\times&j_{\ell}(kr)~j_{\ell}(kr')~\mathrm{Im}~G_{\ell j}^{(P,N)}(r,r';E), \nonumber
 \end{eqnarray}
The single-particle density distribution can be calculated from the hole spectral function in the following way, 
 \begin{equation}
    \label{eq:charge}
    \rho^{(P,N)}(r) = \frac{1}{4\pi} \sum_{\ell j} (2j+1) \int_{-\infty}^{\varepsilon_F}dE\ S^{(P,N)}_{\ell j}(r;E),
 \end{equation}
 where $S^{(P,N)}_{\ell j}(r;E) = S^{(P,N)}_{\ell j}(r,r;E)$ is the diagonal of the hole spectral density.

 The spectral strength for a given $\ell j$ partial wave can be found by summing (integrating) the spectral function according to
 \begin{equation}
    S_{\ell j}^{(P,N)}(E) = \int_0^\infty S_{\ell j}^{(P,N)}(r;E)r^2dr,
    \label{eq:strength}
 \end{equation}
 when $E<\varepsilon_F$.
 The spectral strength $S_{\ell j}^{(P,N)}(E)$ is the contribution at energy $E$ to the occupation from all orbitals with orbital angular momentum $\ell$ and total angular momentum $j$. 
 
While the focus of the present paper is to study high-momentum components in the DOM framework, it is important to note that it is capable to describe the removal of valence protons near the Fermi energy.
Indeed, the DOM self-energy  generates all the ingredients for the calculation of the $(e,e'p)$ cross section for the removal of valence protons, including the relevant distorted wave, and the overlap function with its normalization. The work of Refs.~\cite{Atkinson:2018,Atkinson:2019,Atkinson:2024,Ramon:2026B} demonstrates that the DOM is capable of correctly describing these cross sections with the accompanying spectroscopic factors constrained by other data.
Another important application is the possibility of predicting the neutron skin~\cite{Mahzoon:2017,atkinson2020dispersive,Pruitt:2020,Pruitt:2020C,Atkinson:2024B} or describing it~\cite{calleya2025investigating,Ramon:2026} after an appropriate measurement~\cite{adhikari2022precision}.
However, the smaller neutron skin in Ref.~\cite{calleya2025investigating} generated more high-momentum neutrons than protons in ${}^{48}$Ca providing further motivation for the present study.
\subsection{Energy-dependent geometry and numerical PV integral}
\label{sec:Energy_dependence}
The fully-nonlocal imaginary part of the DOM self-
energy consists of volume, surface, and spin-orbit terms. 
All these terms are factorized into an energy-dependent normalization, a Woods-Saxon form factor (geometry), and a Gaussian nonlocality. Each term is parametrized with the functional forms proposed by Mahaux and Sartor~\cite{Mahaux91}.
In particular, the volume imaginary part of the self-energy is factorized as
\begin{align}
 \label{eq:imnl_main_text}
 W^{vol}_{0\pm}(E) f\left(\tilde{r};r^{vol}_{\pm (p,n)};a^{vol}_{\pm}\right)H \left( \bm{s}; \beta_{\pm (p,n)}^{vol}\right),
 \end{align}
 where $\tilde{r} =(r+r')/2$, $\bm{s}=\bm{r}-\bm{r}'$, $\beta$ is a nonlocality parameter, and $H(\bm{s},\beta)$ is a Gaussian nonlocality and $W^{vol}_{0\pm}(E)$ is defined as
 \begin{equation}
 W^{vol}_{0\pm}(E)\! = \! \Delta W^{\pm}_{NM}(E) + \! 
 \begin{cases}
 0  \\
 A^{vol}_{\pm(p,n)}  \frac{\left(|E-\varepsilon_F|-\mathcal{E}^{vol}_{\pm}\right)^4}
 {\left(|E-\varepsilon_F|-\mathcal{E}^{vol}_{\pm}\right)^4 + (B^{vol}_{\pm})^4} 
 \end{cases}
 \label{eq:volumeS}
 \end{equation}
 where the top line is valid for $\text{if } |E-\varepsilon_F| < \mathcal{E}^{vol}_{\pm}$ and the bottom one for $E-\varepsilon_F| > \mathcal{E}^{vol}_{\pm}$.
A detailed description of the parametrization can be found in App.\ref{app:parametrization}. 
These factorized forms have traditionally been used for the PV integral in Eq.~\eqref{eq:dispersion} to calculate the dispersive correction to the real part from the imaginary part of the self-energy analytically~\cite{vanderkam2000analytical}.
In the present analysis, an energy dependence is introduced for the radius parameter $r_{-(p,n)}^{vol}$ of the volume-imaginary Woods-Saxon form factor, but only in the potential below the Fermi energy,~$\epsilon_{F}$. A linear change of radius as a function of $(E- \epsilon_{F})$,
 \begin{equation}
    r_{-(p,n)}^{vol} (E) = r_{-(p,n)}^{vol} + m_{vol}^{r} (E- \epsilon_F)
    \label{eq:RE}
 \end{equation}
is considered, where $m_{vol}^{r}$ is the slope of the linear change of radius with energy. A positive value of $m_{vol}^{r}$ will therefore decrease the radius as the energy decreases.

\begin{figure}[t]
    \includegraphics[width=\columnwidth]{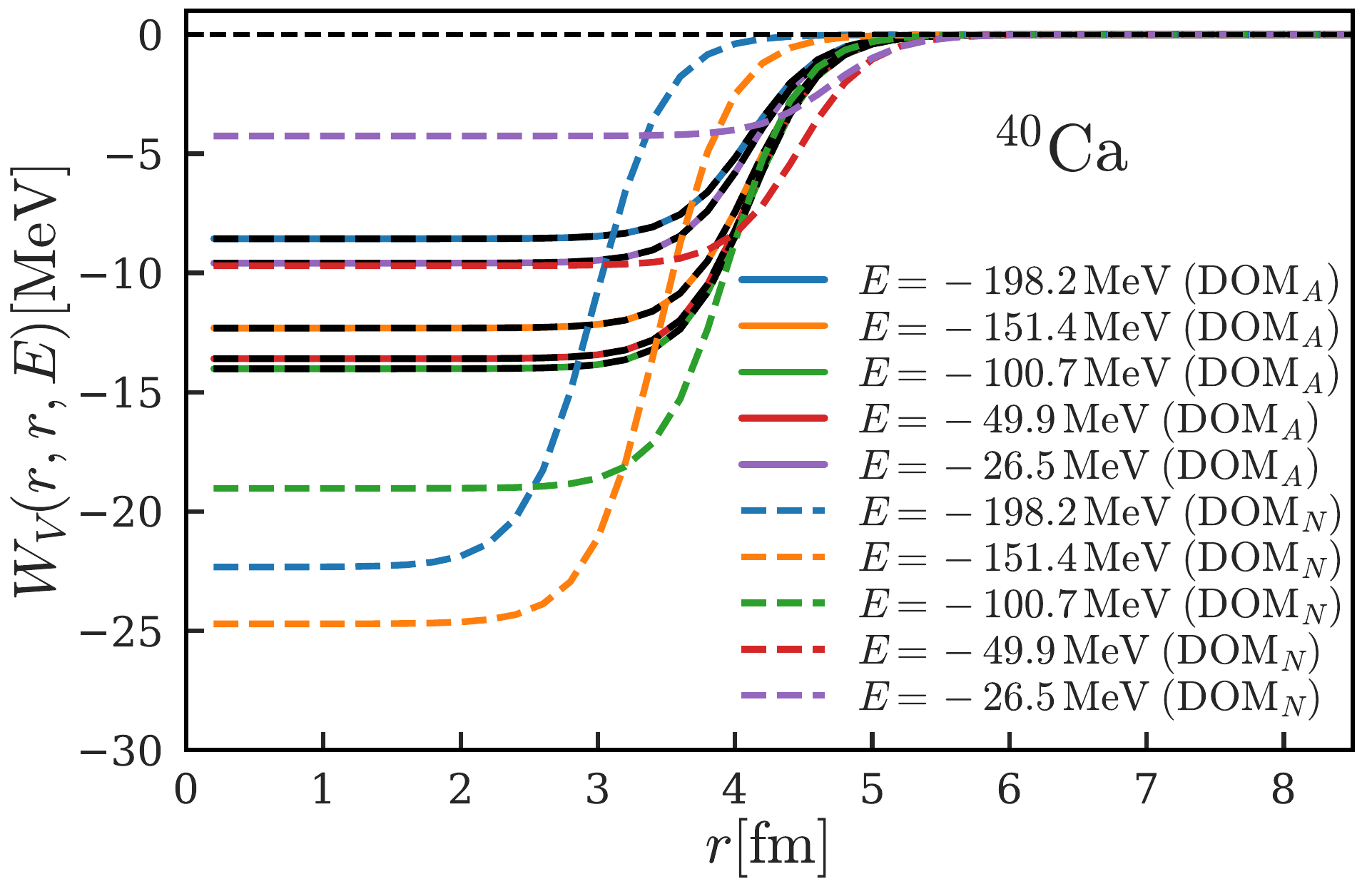}
    \caption{Woods-Saxon geometry of the volume imaginary part of the self-energy without the nonlocality factor. $\mathrm{DOM_A}$ with solid lines indicates the  standard DOM with  $m_{vol}^{r} = m_{vol}^{A}=0$. $\mathrm{DOM_N}$ with dashed lines refers to the geometry with energy dependence in radius, $r_{-(p)}^{vol}$ and depth, $A_{(p)}^{vol}$ implemented by setting $m_{vol}^{r} = 0.01~\mathrm{fm~MeV^{-}}$ and $ m_{vol}^{A} = 0.18~\mathrm{MeV^{-1}}$ in Eq.\eqref{eq:RE} and Eq.\eqref{eq:AE} ${}^{40}$Ca.} 
   \label{fig:vol_img_geometry}
\end{figure}
 Alone, changing the radius with energy would impact the volume integral as well as the bound state properties; \textit{e.g.}, binding energy, position of single particle levels, charge density distribution, and total particle number. To compensate for this decrease in radius with energy, the amplitude, $A_{-(p,n)}^{vol}$ (see the App.\ref{app:parametrization}) was also made energy dependent,
  \begin{equation}
    A_{-(p,n)}^{vol} (E) = A_{-(p,n)}^{vol} - m_{vol}^{A} (E- \epsilon_F).
    \label{eq:AE}
 \end{equation}
Here $m_{vol}^{A}$ is the slope of the linear change of the amplitude with energy.
\begin{table}[b]
\centering
\caption{Parameters used to make the volume imaginary part of the DOM energy dependent.}
\label{tab:slopes}
\begin{ruledtabular}
\renewcommand{\arraystretch}{1.3}
\begin{tabular}{l|cccc}
\multirow{2}{*}{\centering Parameter}& \multicolumn{2}{c}{$^{40}$Ca} & \multicolumn{2}{c}{$^{48}$Ca}\\
\cline{2-3} \cline{4-5}
{}& Proton & Neutron & Proton & Neutron \\
\cline{1-3} \cline{4-5}
$m^{r}_{vol}~[\mathrm{fm~MeV^{-1}}]$ & $0.0100$ & $0.0100$ & $0.00850$ & $0.0135$ \\
$m^{A}_{vol} ~[\mathrm{{MeV}^{-1}}]$ & $0.180$ & $0.180$ & $0.183 $ & $0.170$ \\
\end{tabular}
\end{ruledtabular}
\end{table}
Figure~\ref{fig:vol_img_geometry} depicts Woods-Saxon form factors at different energies, which represent the geometry of the volume-imaginary potential in Eq.~\eqref{eq:volumeS} without the nonlocality factor for ${}^{40}$Ca. Solid lines $(\mathrm{DOM_{A}})$ in Fig.~\ref{fig:vol_img_geometry} shows the shape of the geometry at different energies as a function of radius when $r_{-(p,n)}^{vol}$ and $A_{-(p,n)}^{vol}$ have no energy dependence as generated in Ref.~\cite{Atkinson:2018}. Although all of them have fixed $r_{-(p,n)}^{vol}$, the depth changes according to the energy-dependent part of the potential. Dashed lines $(\mathrm{DOM_{N}})$ represent the geometry with radial energy dependence with $m^{r}_{vol} = 0.010~\mathrm{fm~MeV^{-1}}$ along with the energy-dependent amplitude with $m^{A}_{vol} = 0.180~\mathrm{MeV^{-1}}$. The values of these parameters used in this analysis for ${}^{40}$Ca and ${}^{48}$Ca are collected in TABLE.~\ref{tab:slopes}.

The traditionally used analytical expressions~\cite{vanderkam2000analytical} for the dispersive integral can no longer be used because the integrand cannot be factorized. 
Therefore, we implement a numerical method to evaluate the PV integral using a modified Simpson's rule developed in Ref.~\cite{Amari:1994evaluation}. This numerical method is capable of yielding exact results provided the integrand approaches zero smoothly at the limits of integration. The volume part, $W_{0\pm}^{vol}(E)$, of the imaginary potential goes to zero at a moderately negative energy and at $\epsilon_F$, thus, the PV integration is possible using the numerical method in the negative-energy domain. However, in the positive energy domain, $W_{0\pm}^{vol}(E)$ does not approach zero as originally proposed by Mahaux and Sartor~\cite{Mahaux91}
Since the negative energy domain of the imaginary part contributes to the positive domain of the real part, and vice versa, it is mandatory to perform the PV integration over the entire energy range.
The PV integration for the whole range could be done with the numerical method developed in Ref.~\cite{Amari:1994evaluation} by ensuring that the volume imaginary part smoothly vanishes at some high energy, reflecting the value of the core of a chosen nucleon-nucleon interaction.
This would require a considerable amount of extra computation time, which is not feasible for practical research calculations involving parameter fitting without a clear physical motivation. However, we can combine the analytic PV integral results with numerical results only performed for the negative energy domain, which is computationally less expensive compared to the full range numerical integration. We therefore subtract the numerical result, $N_C$, with a constant radius (and depth) in the negative domain, from the analytical result, $A_{Full}$, for the full domain, and then add the numerical result, $N_E$, with an energy-dependent radius (and depth) for the negative domain only. As shown in the Apps.~\ref{app:bound_state}, \ref{app:scattering}, and \ref{app:params}, the parameters of the other ingredients of the DOM potential only change minimally or not at all to achieve a correct representation of all experimental observables.
Only the description of the high-momentum components has changed.
\begin{figure}[t]
    \includegraphics[width=\columnwidth]{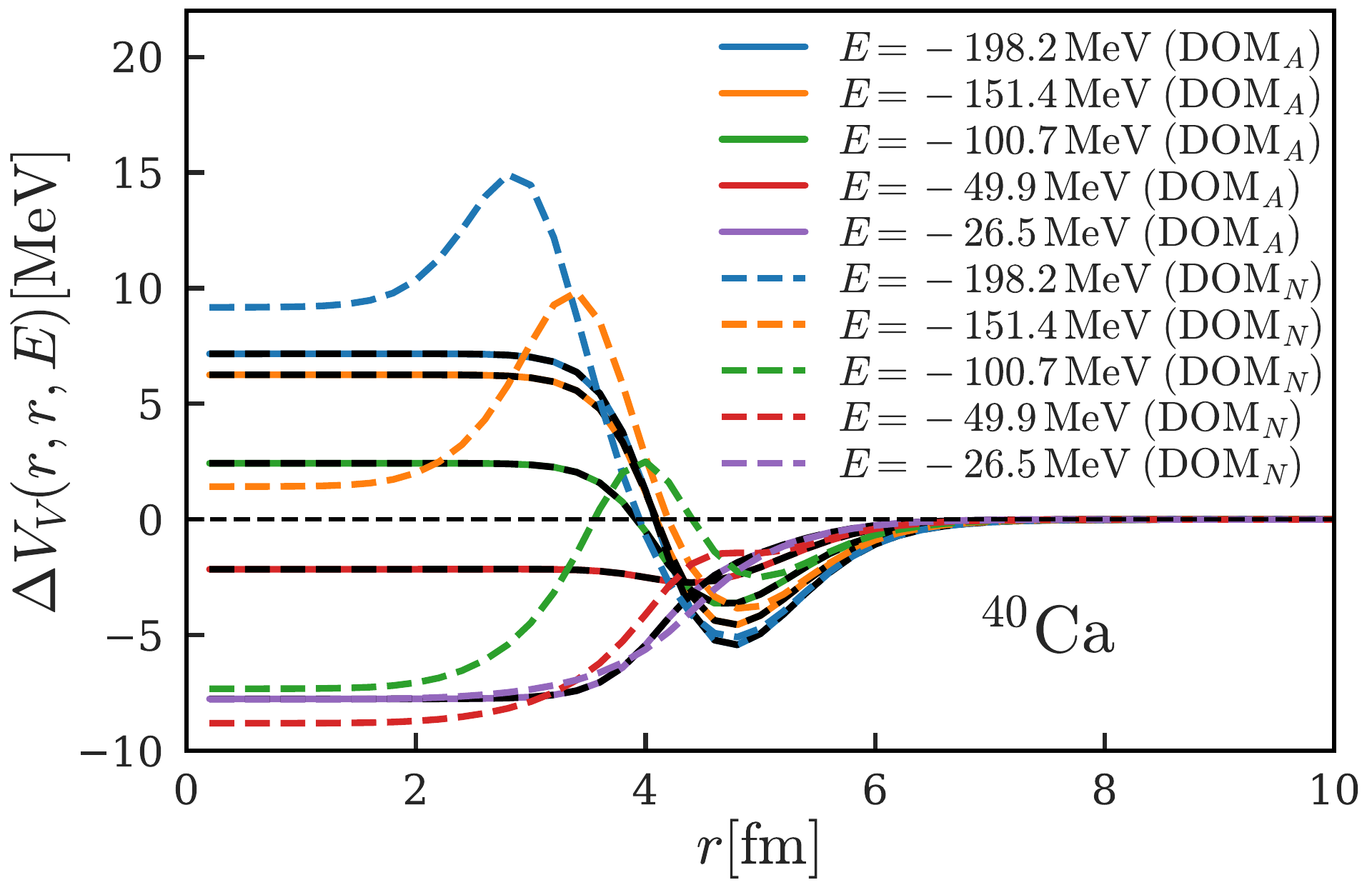}
    \caption{
    Volume dispersive correction to the real part of the DOM self-energy at the diagonal$(r = r')$ for fixed energies utilized in Fig.~\ref{fig:vol_img_geometry}. Colored solid and dashed lines represent the dispersive correction corresponding to the imaginary potential in Fig.~\ref{fig:vol_img_geometry} evaluated with $\mathrm{DOM_A}$ and $\mathrm{DOM_N}$, respectively. Black dashed lines represents $\mathrm{DOM_N}$ results with $m^{r}_{vol} = m^{A}_{vol} = 0$.}
   \label{fig:dis_corr_fixed_E}
\end{figure}

\section{Analytical-numerical equivalence}
\label{sec:equivalence}

\subsection{Volume dispersive correction}
\label{sec:geo_energy_dependence}

\begin{figure}[t]
    \includegraphics[width=\columnwidth]{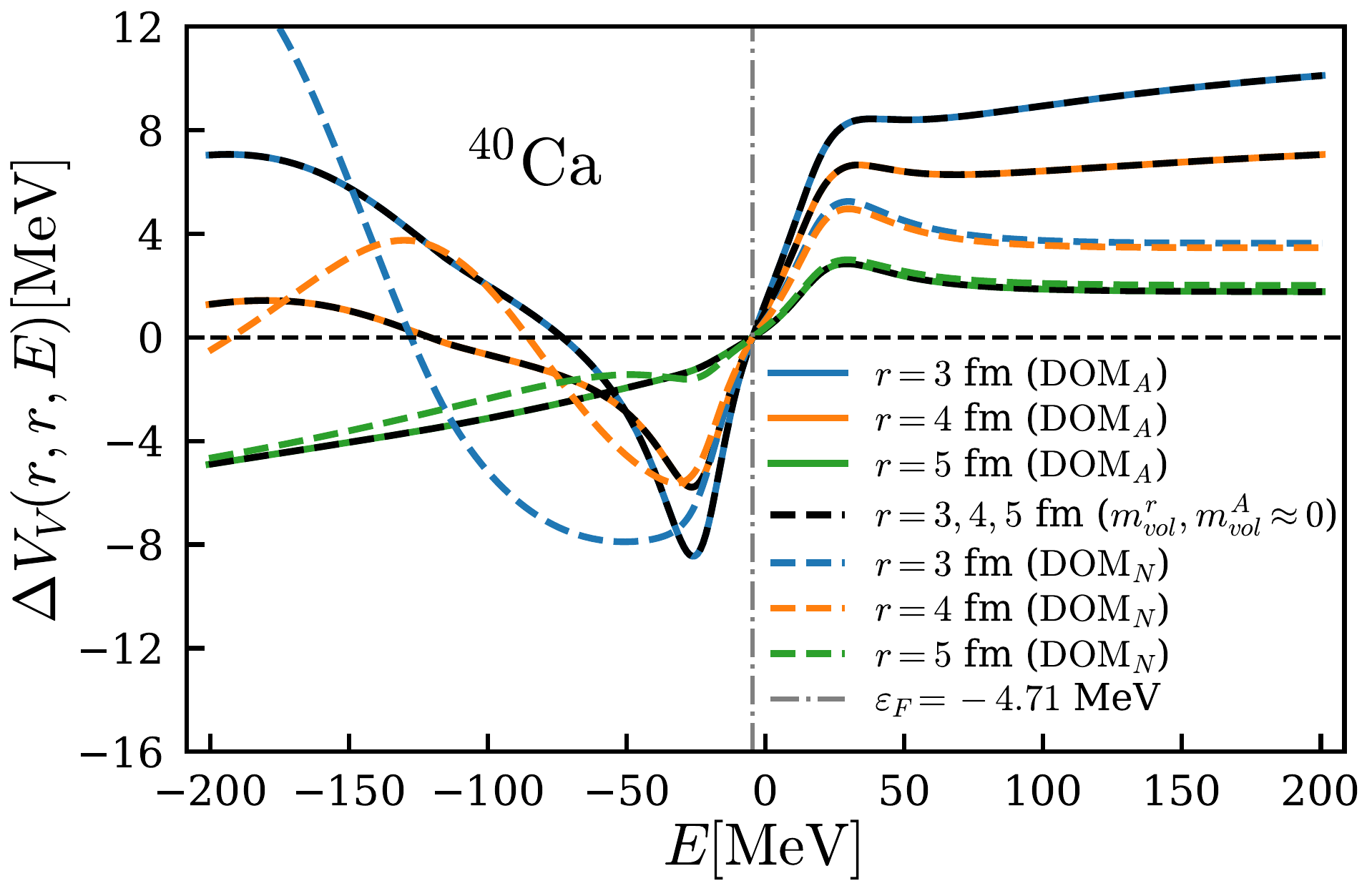}
    \caption{Volume dispersive correction to the real part of the DOM self-energy at the diagonal for fixed $r = r'=3, 4, 5~\mathrm{fm} $. $\mathrm{DOM_A}$ with solid lines indicates the standard DOM. $\mathrm{DOM_N}$ with dashed lines refers to the self-energy with the introduced energy-dependent radius, $r_{-(p)}^{vol}$ and depth, $A_{(p)}^{vol}$. Black dashed lines represents the $\mathrm{DOM_N}$ results with $m^{r}_{vol} = m^{A}_{vol} = 0$.}
   \label{fig:dis_corr_fixed_r}
\end{figure}

It is mandatory to benchmark this numerical extension of the dispersive optical model $(\mathrm{DOM_N})$ against the analytical version $(\mathrm{DOM_A})$ with energy-independent $r_{-(p,n)}^{vol}$ and $A_{-(p,n)}^{vol}$. 
We plot the dispersive correction $\Delta V_{V}(r,r';E)$ using two different perspectives: one with fixed energy (Fig.~\ref{fig:dis_corr_fixed_E}) and another with fixed radius (Fig.~\ref{fig:dis_corr_fixed_r}) for both analytical and numerical results.

Figure~\ref{fig:dis_corr_fixed_E} shows diagonal elements ($r=r'$) of the dispersive correction $\Delta V_{V}(r,r;E)$ for fixed energies ranging from -26.5 to 198.2~MeV. The solid and dashed lines with colors are the results of the dispersive correction to the real part corresponding to the volume-imaginary potential, $W_{V}(r,r';E)$ shown in Fig.~\ref{fig:vol_img_geometry}. Solid colored lines represent analytic results ($\mathrm{DOM_A}$) with $r_{-(p,n)}^{vol}$ and $A_{-(p,n)}^{vol}$, while dashed colored lines represent numerical results $(\mathrm{DOM_N})$ with $r_{-(p,n)}^{vol}(E)$ and $A_{-(p,n)}^{vol}(E)$ with $m^{r}_{vol} = 0.01~\mathrm{fm~MeV^{-1}}$ and $m^{A}_{vol}=0.18~\mathrm{MeV^{-1}}$. The black dashed lines in Fig.~\ref{fig:dis_corr_fixed_E} correspond to $
\mathrm{DOM_N}$ results with $m^{r}_{vol} = m^{r}_{vol} = 0$, which vanishes the energy-dependent terms in Eq.\eqref{eq:RE} and Eq.\eqref{eq:AE}, and the numerical results perfectly overlap with the analytic results.
This verifies the validity of the new numerical method.

Figure~\ref{fig:dis_corr_fixed_r} displays the dispersive corrections $\Delta V_{V}(r,r';E)$, corresponding to the volume imaginary potential as a function of energy for some selected $r$-values on the diagonal $(r=r')$. Similar to Fig.~\ref{fig:dis_corr_fixed_E}, the solid colored lines $(\mathrm{DOM_{A}})$ and the dashed black lines $(\mathrm{DOM_{N}})$ coincide with each other, again demonstrating their equivalence. It should be noted that the curves of $\Delta V_{V}(r, r';E)$ with a smaller radius are more affected by the new radial energy dependence than the curves with a larger radius, emphasizing the change in the absorption mechanism in the interior introduced here.

\subsection{Bound state properties}
\label{sec:bound_state_properties}

The energy-dependent geometry is introduced only for the volume-imaginary part of the self-energy in the negative energy domain. Hence, it is important to verify that this extension to the DOM preserves the ability to describe ground-state experimental observables.
For ${}^{40}$Ca, the $\mathrm{DOM_{A}}$ and $\mathrm{DOM_{N}}$ calculated total binding energies are $ E_0^\mathrm{DOM_{A}} = 341.03$~MeV and  $E_0^\mathrm{DOM_{N}} = 346.98$~MeV, respectively, which are both close to the experimental value of $E_0^\mathrm{Exp} = 342.05$~MeV~\cite{wang2021ame}. Similarly, the RMS charge radii also have values $r^\mathrm{RMS}_\mathrm{DOM_{A}} = 3.484$~fm and $r^\mathrm{RMS}_\mathrm{DOM_{N}} = 3.491$~fm, compared to the experimental value of $r^\mathrm{RMS}_\mathrm{Exp} = 3.4776$~fm~\cite{Angeli:2013}.
Figure~\ref{fig:chd} shows excellent agreement with the experimental charge density calculated with the standard dispersive optical model, $\mathrm{DOM_A}$, and with the present energy-dependent geometry extension, $\mathrm{DOM_{N}}$.
\begin{figure}[b]
    \includegraphics[width=\columnwidth]{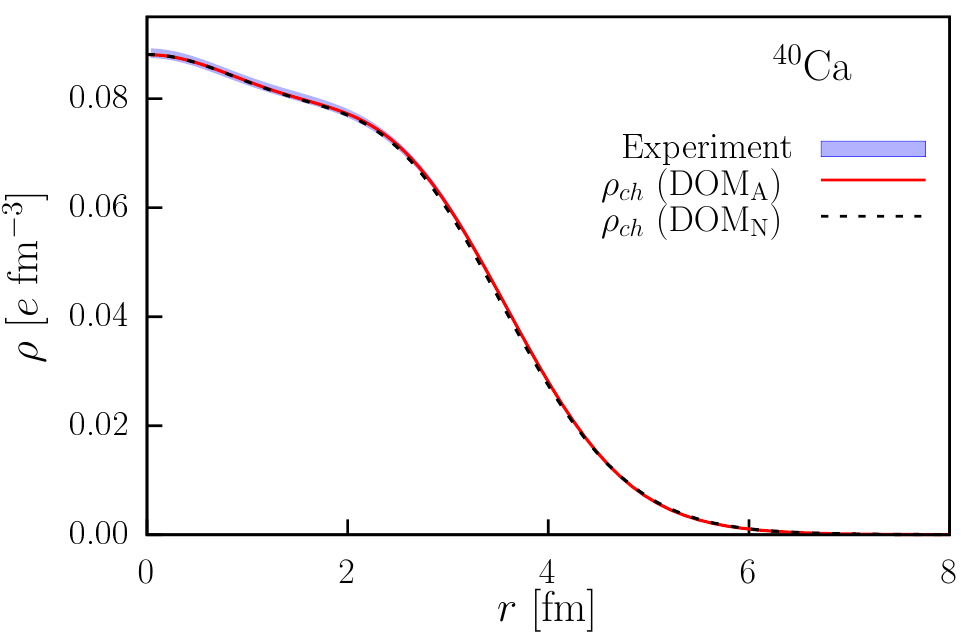}
    \caption{Experimental charge density of $^{40}$Ca~\cite{deVries:1987,Sick79} compared with the DOM-calculated charge density. $\mathrm{DOM_A}$ represents the charge density calculated when the geometry of the volume imaginary has no energy dependence. $\mathrm{DOM_N}$  shows a similar quality of charge density distribution with an energy-dependent geometry of the volume imaginary part of the self-energy.}
   \label{fig:chd}
\end{figure}
Table~\ref{tab:levels} compares the single-particle levels, and Table~\ref{tab:spectroscopic} includes the valence spectroscopic factors for ${}^{40}$Ca calculated with both DOM versions.
\begin{table}[ht]
\centering
\caption{Comparison of proton and neutron single-particle levels in $^{40}$Ca. Here $\mathrm{DOM_A}$ represents the levels calculated using the analytical version of DOM, and $\mathrm{DOM_N}$ indicates levels calculated with the energy-dependent geometry.}
\label{tab:levels}
\begin{ruledtabular}
\renewcommand{\arraystretch}{1.3}
\begin{tabular}{l|cccccc}

& \multicolumn{3}{c}{Proton levels [MeV]} & \multicolumn{3}{c}{Neutron levels [MeV]} \\

\cline{2-4} \cline{5-7}

$n\ell j$ & Exp~\cite{Mueller:2011}. & $\mathrm{DOM_A}$ & $\mathrm{DOM_N}$ & Exp~\cite{Mueller:2011}. & $\mathrm{DOM_A}$ & $\mathrm{DOM_N}$ \\

\cline{1-4} \cline{5-7}
$0d\frac{5}{2}$ & $-14.30$ & $-14.81$ & $-15.38$ & $-22.26$ & $-22.10$ & $-22.69$ \\
$1s\frac{1}{2}$ & $-10.80$ & $-9.84$ & $-9.93 $ & $-18.29$ & $-17.11$ & $-17.29 $ \\
$0d\frac{3}{2}$ & $-8.30$ & $-8.66$ & $-8.82$ & $-15.64$ & $-15.79$ & $-15.94$ \\
$0f\frac{7}{2}$ & $-1.09$ & $-2.99$ & $-3.00 $ & $-8.36$ & $-9.90$ & $-9.90$\\
$1p\frac{3}{2}$ & $-$ & $-$ &$-$ & $-5.86$ & $-6.42$ & $-6.28 $ \\
$1p\frac{1}{2}$ & $-$ & $-$ & $-$& $-4.20$ & $-4.25$ & $-4.07$\\
$0f\frac{5}{2}$ & $-$ & $-$ & $-$& $-1.38$ & $-4.19$ &$-4.05 $ \\
\end{tabular}
\end{ruledtabular}
\end{table}
 The experimentally extracted Nikhef results are also included, and it should be noted that the DOM results generate an excellent agreement with the experimental $(e,e'p)$ cross sections~\cite{Atkinson:2018}. Similar results are obtained for $^{48}$Ca and are displayed in the App.\ref{app:bound_state48Ca}.
\begin{table}[h]
\caption{Comparison of $^{40}$Ca spectroscopic factors extracted by the Nikhef analysis~\cite{Kramer:1989} to the normalization of the corresponding overlap functions obtained in the DOM calculation.}
\label{tab:spectroscopic}
\begin{ruledtabular}
\begin{tabular}{cccc}
\multirow{2}{*}{\centering $n\ell j$} & \multicolumn{3}{c}{\textrm{\textrm{$\mathcal{Z}^{n}_{\ell j}$}}}\\
 \noalign{\vskip 0.1cm}
 \cline{2-4}
 \noalign{\vskip 0.1cm}
 & $\mathrm{DOM_A}$& $\mathrm{DOM_N}$ & \textrm{Ref.}~\cite{Kramer:1989}\\
\colrule
 \noalign{\vskip 0.1cm}
 \vspace{0.1cm}
$0d{\frac{3}{2}}$& $0.67$ & $0.69$ & $0.65$\\
 \vspace{0.1cm}
$1s{\frac{1}{2}}$ & $0.71$ & $0.73$ & $0.51$\\

\end{tabular}
\end{ruledtabular}
\end{table}

These results demonstrate that incorporating energy dependence in the geometry of the volume-imaginary part below the Fermi level does not limit the ability of the DOM to reproduce the same results as the conventional DOM and the available experimental values.
The positive energy domain only faces a small contribution from this extension through the dispersion relation, Eq.~\eqref{eq:dispersion}, corresponding to the volume-imaginary part in the negative energy domain. We therefore present the positive energy results only in App.\ref{app:scattering}.

\section{Volume Integrals and Spectral Functions}
\label{sec:volint}
\begin{figure}[t]
    \includegraphics[width=\columnwidth]{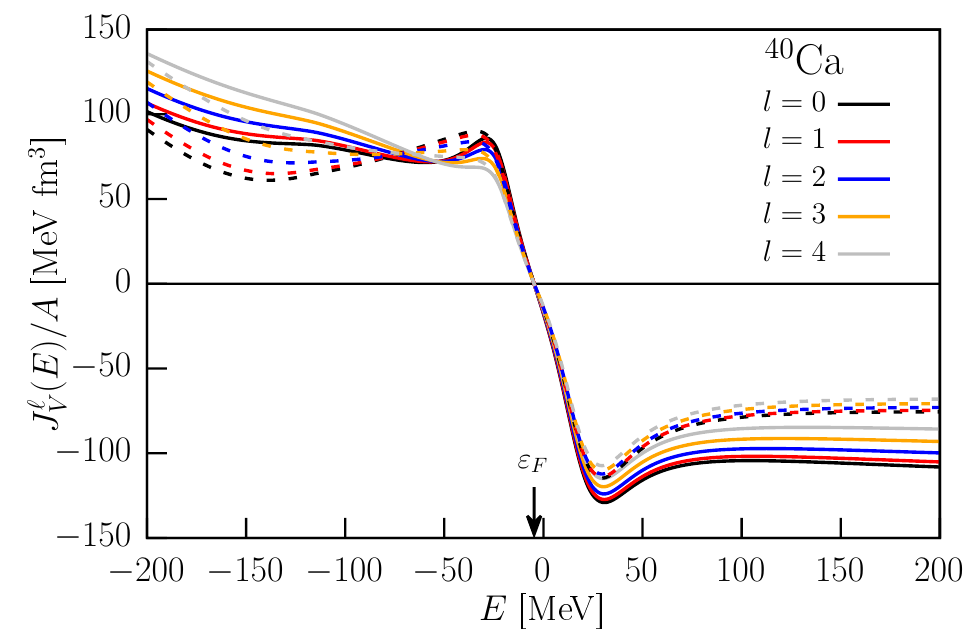}
    \caption{Volume integrals of $\Delta V_V(r,r';E)$ as a function of energy for different $\ell$-values. Solid and dashed lines represent the calculated volume integrals with $\mathrm{DOM_{A}}$ and $\mathrm{DOM_{N}}$, respectively.}
   \label{fig:volint_dis_corr}
\end{figure}

To visualize the effects of the introduced energy-dependent geometry, volume integrals of the volume dispersive correction, defined as
\begin{equation}
    J_{V}^{\ell}(E)/A = \frac{4 \pi}{A} \int dr r^{2} \int dr' r'^{2} \Delta V_V(r,r'E),
    \label{eq:volint}
\end{equation}
are displayed in Fig.~\ref{fig:volint_dis_corr}.
The volume integral is a good quantity to monitor the effect of the DOM extension to $\mathrm{DOM_N}$ as it compresses the full spatial dependence of the self-energy into a single energy-dependent quantity for each $ \ell$-value. 

In Fig.~\ref{fig:volint_dis_corr}, solid lines represent the volume integral of $\Delta V_V(r,r';E)$ with constant geometry $(\mathrm{DOM_A})$ 
for different angular momenta, $\ell$, most relevant in this energy domain. The dashed lines corresponding to the calculations done with $\mathrm{DOM_N}$. Energy dependence in $r_{-(p,n)}^{vol}$ should reduce $J_V^\ell(E)$ (the dashed lines in Fig.~\ref{fig:volint_dis_corr}) drastically as the radius is decreasing with energy; however, $A_{-(p,n)}^{vol}(E)$ increases with energy to compensate, keeping the volume integral close to the analytical results. This energy dependence in $r_{-(p,n)}^{vol}(E)$ and $A_{-(p,n)}^{vol}(E)$ is redistributing the contributions to the volume integral instead of totally changing it. It should be noted that the change in the negative energy domain is contributing to the positive energy domain, even though it is modest.

\begin{figure}[b]
    \includegraphics[width=\columnwidth]{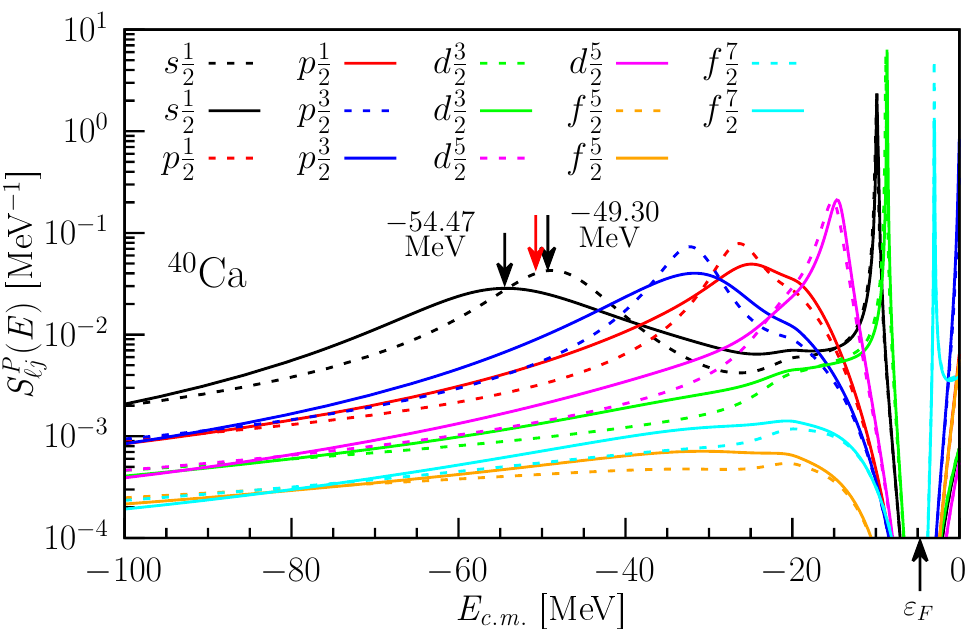}
    \caption{Proton spectral strength of $^{40}$Ca. Solid and dashed lines represent the calculated strength with $\mathrm{DOM_{A}}$ and $\mathrm{DOM_{N}}$, respectively. The black arrows indicate the $0s\frac{1}{2}$ peak obtained with $\mathrm{DOM_A}(-54.47~\mathrm{MeV})$ and $\mathrm{DOM_N}(-49.30~\mathrm{MeV})$. The red arrow indicates the peak location of $0s\frac{1}{2}$ at -50.74~MeV obtained from quasi-free $(p,2p)$ scattering~\cite{jacob1973quasi}.}
   \label{fig:spectral_p_40Ca}
\end{figure}
\begin{figure}[t]
    \includegraphics[width=\columnwidth]{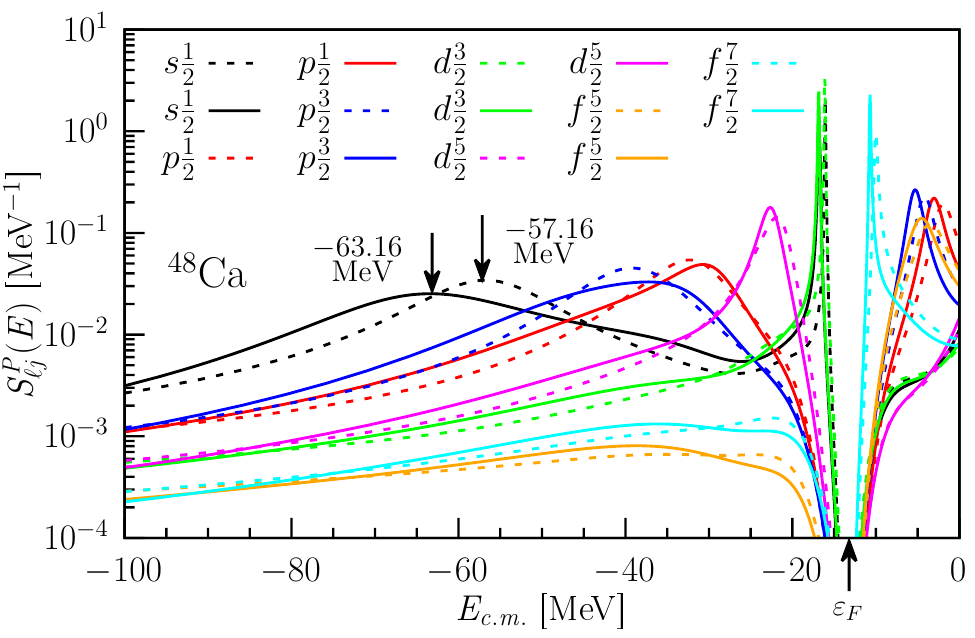}
    \caption{Proton spectral strength of $^{48}$Ca. Solid and dashed lines represent the calculated results with $\mathrm{DOM_{A}}$ and $\mathrm{DOM_{N}}$, respectively. The arrows indicate the $0s\frac{1}{2}$ peak obtained with $\mathrm{DOM_A}(-63.16~\mathrm{MeV})$ and $\mathrm{DOM_N}(-57.16~\mathrm{MeV})$.}
   \label{fig:spectral_p_48Ca}
\end{figure}

Spectral strengths according to Eq.~\eqref{eq:strength} calculated with $\mathrm{DOM_A}$ (dashed lines) and $\mathrm{DOM_N}$ (solid lines) are displayed in Fig.~\ref{fig:spectral_p_40Ca} and Fig.~\ref{fig:spectral_p_48Ca} for $^{40}$Ca and $^{48}$Ca, respectively.
For the latter Ca isotope similar results between the different DOM implementations are obtained. We focus here on those quantities below the Fermi energy where different results are generated.
The values of the newly-introduced parameters used in Eq.~\eqref{eq:RE} and Eq.~\eqref{eq:AE} are shown in Table.~\ref{tab:slopes}.
The correspondence between the $\mathrm{DOM_A}$ and $\mathrm{DOM_N}$ results displayed in Fig.~\ref{fig:spectral_p_48Ca} is comparable to the correspondence between volume integrals in Fig.~\ref{fig:volint_dis_corr}, which is inevitable given that the bound state properties of both methods match so well (see Sec.~\ref{sec:bound_state_properties}).   
Figure~\ref{fig:volint_dis_corr} shows that the $\mathrm{DOM_A}$ and $\mathrm{DOM_N}$ generate similar volume integrals from the Fermi energy ($-13.21$~MeV for $^{40}$Ca proton) to around $-20$~MeV for $\ell=0$ and  to around $-30$~MeV for $\ell=4$, while the volume integrals are slightly different below $-30$~MeV. This same feature is observed in spectral strength plots (Fig.~\ref{fig:spectral_p_40Ca} and Fig.~\ref{fig:spectral_p_48Ca}). The sharp spectral function peaks for valence hole states just below the Fermi energy are minimally affected, while the deeper bound state peaks are strongly modified. The $0s\frac{1}{2}$ state is affected the most, with its peak location moving from $-54.47 \pm 13.20$~MeV to  $-49.30 \pm 7.1$~MeV for $^{40}$Ca and from $-63.17 \pm 16.10$~MeV to $-57.16 \pm 9.50$~MeV for $^{48}$Ca. These energies are indicated in the spectral function plots (See Fig.\ref{fig:spectral_p_40Ca} and Fig.\ref{fig:spectral_p_48Ca}) with arrows, and uncertainties are half of the FWHM of the corresponding peaks.  
From quasi-free $(p,2p)$ scattering experiments, the $0s\frac{1}{2}$ peak location is $-50.74 \pm 5.66~\mathrm{MeV}$~\cite{jacob1973quasi} for $^{40}$Ca. The $\mathrm{DOM_N}$ result of $-49.30 \pm 7.1$~MeV represents a clear improvement over the $\mathrm{DOM_A}$ value of $-54.47 \pm 13.20$~MeV, indicating a better description of the self-energy for deeply bound levels.
This further confirms the general notion that an energy-dependent geometry generates an improved description of the properties of deeply bound nucleons and as shown in the next section, the spectral properties of high-momentum components.

\section{Results for high-momentum components}
\label{sec:results}

\begin{figure}[b]
    \includegraphics[width=\columnwidth]{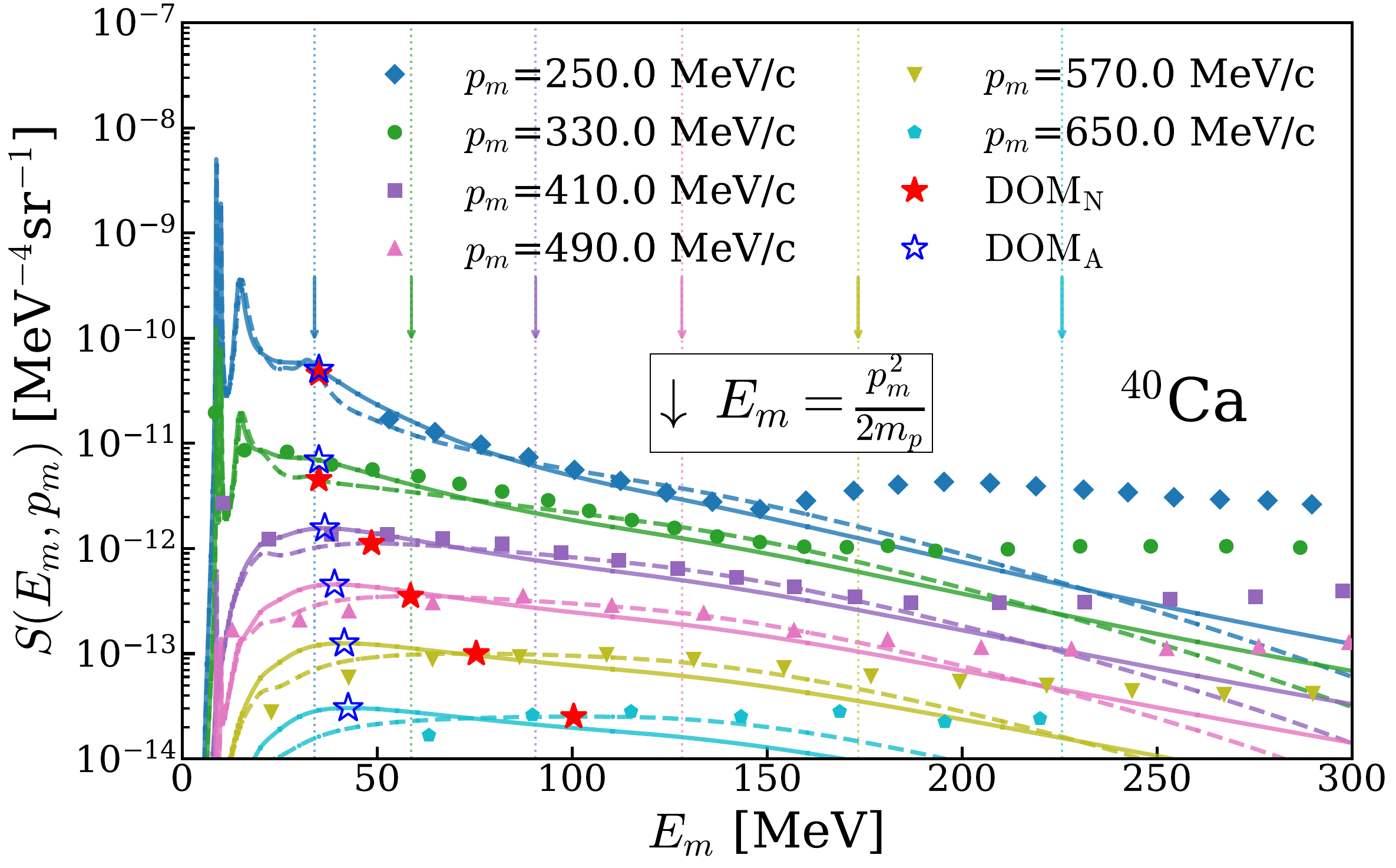}
    \caption{Proton spectral function as a function of missing energy for different missing momenta as indicated in the figure. Solid and dashed lines represent calculated spectral functions with $\mathrm{DOM_{A}}$ and $\mathrm{DOM_{N}}$, respectively. Blue and red stars on the solid and dashed lines represent the maxima of $\mathrm{DOM_{A}}$ and $\mathrm{DOM_{N}}$ calculated spectral functions, respectively. The maxima selection range was set to avoid the single-particle region. The data are the average of the $^{27}$Al and $^{56}$Fe measurements from Ref.~\cite{Rohe:2004thesis}.} 
   \label{fig:Spectral_Strength_40Ca}
\end{figure}

The effect of the newly introduced energy-dependent geometry can be clearly demonstrated in the proton spectral function, shown in Fig.~\ref{fig:Spectral_Strength_40Ca}.
In Fig.~\ref{fig:Spectral_Strength_40Ca}, the proton spectral function of $^{40}$Ca is plotted as a function of missing energy, $E_{m}$, and missing momentum, $p_{m}$. 
Dashed lines represent the spectral function calculated using $\mathrm{DOM_A}$ while solid lines represent the spectral function calculated using $\mathrm{DOM_N}$ (with an energy-dependent geometry). 
While there is no corresponding experimental data for $^{40}$Ca, we include data points which are an average of $^{27}$Al and $^{56}$Fe proton knockout measurements~\cite{Rohe:2004thesis} as was done in Ref.~\cite{Mahzoon:2014}. 

\begin{figure}[b]
    \includegraphics[width=\columnwidth]{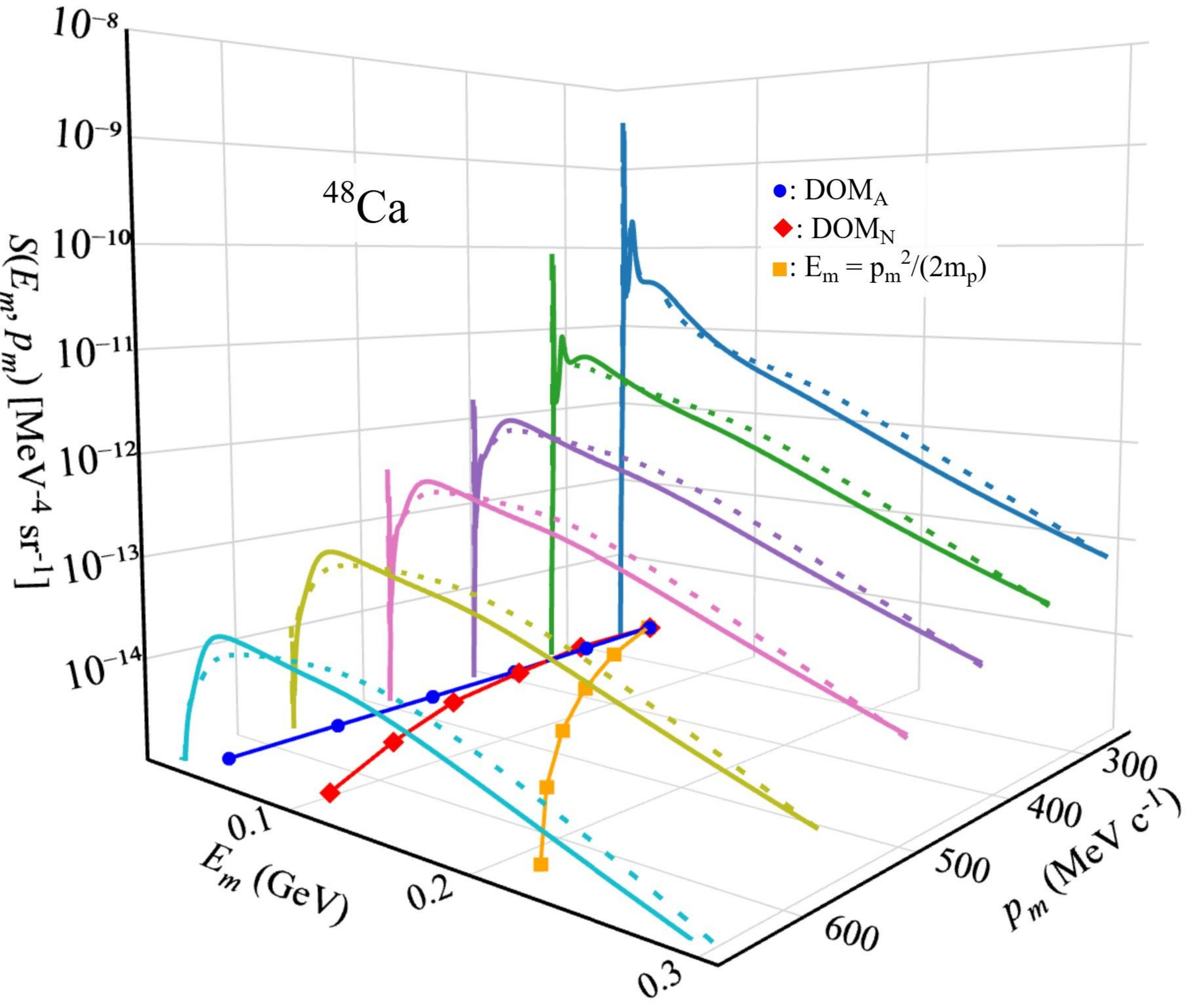}
    \caption{Proton spectral functions in $^{48}$Ca as a function of missing energy for different missing momenta as indicated in Fig.~\ref{fig:Spectral_Strength_40Ca}. Solid and dashed lines represent calculated spectral functions with $\mathrm{DOM_{A}}$ and $\mathrm{DOM_{N}}$, respectively. 
    Blue circles and red diamonds represent the maxima of $\mathrm{DOM_{A}}$ and $\mathrm{DOM_{N}}$ calculated spectral strengths for $^{48}$Ca, respectively. Orange squares correspond to the location of $(E_m = p_{m}^2/{2m_p}, p_m)$.
    }
   \label{fig:Spectral_Strength_48Ca}
\end{figure}

The maximum of each empirical data set in Fig.~\ref{fig:Spectral_Strength_40Ca} increases in $E_m$ as $p_m$ increases. 
The maximum of the $p_m=250~\mathrm{MeV/c}$ data set (diamonds) occurs at roughly $E_m = 40$~MeV while the maximum of the $p_m=650~\mathrm{MeV/c}$ data set (pentagons) occurs slightly above $E_m = 100$~MeV.  
 This behavior is not captured in the old $\mathrm{DOM_{A}}$ implementation; the location of the corresponding maxima in $E_m$ (indicated by the hollow blue stars on the dashed lines in Fig.~\ref{fig:Spectral_Strength_40Ca}) barely changes between curves of varying $p_m$.
 However, when energy dependence is introduced in the geometry of the potential, the corresponding maximum point $E_m$ (indicated by solid red stars on the solid lines) noticeably increases as $p_m$ grows. 
 This is the exact behavior we hoped to achieve with the energy-dependent geometry, and this is because it simulates qualitatively the \textit{ab initio} behavior of the self-energy. We point to Fig.~2 of Ref.~\cite{Muther:1994} for the $p\frac{1}{2}$ spectral function of ${}^{16}$O and Fig.~2 of Ref.~\cite{Dussan:2011} for the $d\frac{3}{2}$ spectral function for ${}^{40}$Ca to illustrate this behavior.
 Such theoretical calculations observe simultaneous energy and momentum conservation and the current DOM implementation appears to capture this properly.
 Just as with the experimental observation, there is a notable deviation from the naive expectation that the peaks will occur at $E_m = p_m^2/2m$ (indicated by arrows in Fig.~\ref{fig:Spectral_Strength_40Ca}. 
 This is not surprising as this is based on a free particle energy in the intermediate state of the self-energy which should be replaced by a spectral distribution that includes a broadening and energy shift.
A similar result incorporating the energy-dependent geometry is seen for $^{48}$Ca, as demonstrated by the spectral functions plotted in Fig.~\ref{fig:Spectral_Strength_48Ca}. 
In this alternate perspective, we track the maximum of $S(E_m,p_m)$ with circles for $\mathrm{DOM_A}$ and diamonds for $\mathrm{DOM_N}$, while the squares represent the locations of $E_m=p_m^2/2m_p$.
The introduction of the energy-dependent geometry succeeds in generating more realistic properties of high-momentum spectral functions.

\begin{figure}[t]
    \includegraphics[width=\columnwidth]{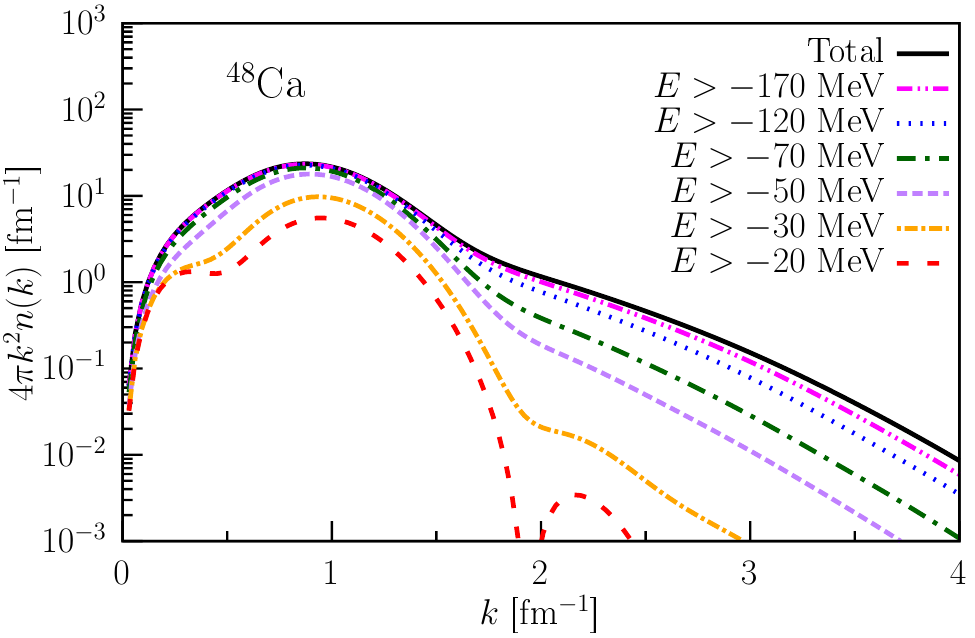}
    \caption{Proton momentum distributions of $^{48}$Ca, obtained with different energy cut-offs in the integration of the spectral functions calculated with $\mathrm{DOM_N}$.}
   \label{fig:kdist_Ecutoff}
\end{figure}

To further investigate the connection between missing momentum and missing energy, we consider momentum distributions (see Fig.~\ref{fig:kdist_Ecutoff}) obtained by integrating spectral functions over missing energy, 
 \begin{equation}
    \label{eq:momentum_distribution}
    n^{(P,N)}(k) =\sum_{\ell j} (2j+1) \int_{-\infty}^{\varepsilon_F}dE\ S^{(P,N)}_{\ell j}(k;E).
 \end{equation}
The solid line in Fig.~\ref{fig:kdist_Ecutoff} represents the integral in Eq.~\eqref{eq:momentum_distribution} over the full energy domain for quasi-holes (where we use -200~MeV as the lower bound in Eq.~\eqref{eq:momentum_distribution} since the spectral functions are negligible beyond -200~MeV, as seen in~\cref{fig:Spectral_Strength_40Ca,fig:Spectral_Strength_48Ca}). Also included in Fig.~\ref{fig:kdist_Ecutoff} are distributions generated from partial energy intervals with varying cutoffs (replacing the lower bound of the integral in Eq.~\eqref{eq:momentum_distribution} with the energies listed in the figure legend).

As the energy cutoff increases from -200~MeV to -70~MeV, the reduction of $n(k)$ in Fig.~\ref{fig:kdist_Ecutoff} primarily occurs for momenta $k> 1.52$ fm$^{-1}$ (300~MeV/c). This region of $n(k)$ is considered the high momentum region. The change between a cutoff of -70~MeV and -50~MeV now visually reduces $n(k)$ not only for high momenta, but also for low momenta. This is expected, since increasing the energy cutoff to -50~MeV now excludes a large fraction of the $0s\frac{1}{2}$ orbital, which is evident by the fact that the broad peak of the $s\frac{1}{2}$ spectral strength in Fig.~\ref{fig:spectral_p_48Ca} has considerable strength below the -50~MeV cutoff. The impact on lower momenta becomes more substantial as the cutoff is further increased, not only from the exclusion of more peaks from Fig.~\ref{fig:spectral_p_48Ca}, but also because the excluded peaks are narrower and correspond to the valence $1s\frac{1}{2}$ and $0d\frac{3}{2}$ orbits. 

Indeed the shortest energy interval, with a cutoff of -20~MeV in Fig.~\ref{fig:kdist_Ecutoff}, includes only these sharp ``delta-like" peaks corresponding to the valence $0d\frac{3}{2}$ and $1s\frac{1}{2}$ orbitals. These orbitals are the closest to being described in an IPM because the imaginary part of the self-energy is negligible in the region near $\varepsilon_F$. 
The strength of the imaginary volume integral, $J_\ell^W(E)$, directly determines the fragmentation of strength in $S(E)$ (or the width of the curves in Fig.~\ref{fig:Spectral_Strength_48Ca}), which is caused by underlying many-body correlations (which include short-range correlations). Thus, the sharp peaks near the Fermi energy are not highly correlated, so it is not 
surprising that the momentum distribution corresponding to the \mbox{-20}~MeV cutoff containing only these peaks in Fig.~\ref{fig:kdist_Ecutoff} has very little strength in the high momentum region as confirmed by $(e,e'p)$ cross sections~\cite{Atkinson:2018}. Similarly, this is why the lower energy cutoffs, which include the continuum part of the spectral function built on many-body correlations, exhibit the most high-momentum content.
Figure~\ref{fig:Spectral_Strength_48Ca} also confirms the tendency of finding the high-momentum strength at large missing energy predicted in Fig. 2 of Refs. ~\cite{Muther:1994} for ${}^{16}$O, and in Fig.2 of Ref~\cite{Dussan:2011} for ${}^{40}$Ca.

We quantify the high-momentum content by calculating the percentage of the momentum distribution with $k>1.52$ fm$^{-1}$. As seen in Table~\ref{tab:high_momentum_Ecutoffs}, the high-momentum fraction increases by over a factor of 5 from 1.44\% for the highest energy cutoff to 8.18\% for the entire quasi-hole energy domain. 
\begin{table}[t]
\centering
\caption{Proton high-momentum fraction in $^{48}$Ca calculated with $\mathrm{DOM_{N}}$ for different energy cut-offs. Momenta above $k >1.52~\mathrm{fm^{-1}}~(300~\mathrm{MeV/c}) $ are considered as high-momentum\cite{Hen:2014momentum, Duer:2018, Duer:2019direct}.} 
\label{tab:high_momentum_Ecutoffs}
\begin{ruledtabular}
\renewcommand{\arraystretch}{1.2}
\begin{tabular}{c|c}
Energy cut-offs [MeV] & High-momentum fraction [\%] \\
\cline{1-2}
$-20$ & $1.44$\\
$-30$ & $2.00$\\
$-50$ & $3.21$\\
$-70$ & $4.12$ \\
$-120$ & $6.16$ \\
$-170$ & $7.40$ \\
Total & $8.18$ \\
\end{tabular}
\end{ruledtabular}
\end{table}
From the results in Fig.\ref{fig:kdist_Ecutoff} and Table~\ref{tab:high_momentum_Ecutoffs}, it is evident that an increase in the momentum content of the single-particle strength occurs with increasing excitation energy in the $A-1$ system~\cite{polls1995short,Exposed!}. 
 This confirms the interpretation that energy and momentum conservation will dictate that high-momentum particles are more likely to occur at higher excitation energies. 
 This reflects the trend revealed by the empirical data and the new $\mathrm{DOM_N}$ spectral functions in~\cref{fig:Spectral_Strength_40Ca,fig:Spectral_Strength_48Ca} where increasing $p_m$ corresponds to the maximum of $S(p_m,E_m)$ increasing in $E_m$. 
 The same feature is observed in nuclear matter, where the peak of the single-particle spectral function for momenta above the Fermi momentum, $k_F$, increases in energy as $k^{2}$~\cite{RAMOS19891,BENHAR1989267,Ciofi:1991}.

 The close connection between $n(k)$ and $S(k,E)$ dictates that the improved description of the spectral functions, $S(p,E)$, in~\cref{fig:Spectral_Strength_40Ca,fig:Spectral_Strength_48Ca} leads to an improved description of the momentum distribution. 
 \begin{figure}[t]
    \includegraphics[width=\columnwidth]{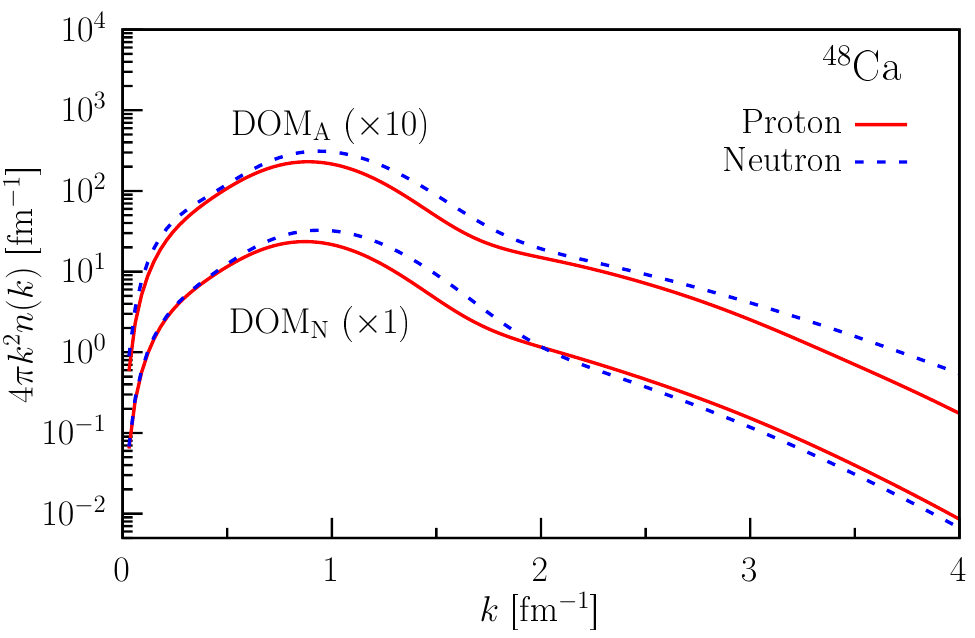}
    \caption{Comparison of $^{48}$Ca momentum distributions calculated with $\mathrm{DOM_A}$~\cite{calleya2025investigating} (top curves) and $\mathrm{DOM_N}$ (bottom curves). Solid lines indicate protons and dashed lines indicate neutrons.}
   \label{fig:kdist_48Ca_A_N}
\end{figure}
\begin{table}[h]
\caption{Comparison of nucleon high-momentum fraction in $^{48}$Ca obtained using $\mathrm{DOM_{A}}$~\cite{calleya2025investigating} and $\mathrm{DOM_{N}}$. Momentum above $k >1.52~\mathrm{fm^{-1}}~(300~\mathrm{MeV/c)} $ are considered as high-momentum\cite{Hen:2014momentum, Duer:2018, Duer:2019direct}.}
\label{tab:high_momentum_Ca48_A_N}
\begin{ruledtabular}
\begin{tabular}{ccc}
\multirow{2}{*}{Nucleon} &
\multicolumn{2}{c}{High-momentum fraction [\%]}\\
 \noalign{\vskip 0.1cm}
\cline{2-3}
 \noalign{\vskip 0.1cm}
& $\mathrm{DOM_A}$ & $\mathrm{DOM_N}$ \\
\colrule
 \noalign{\vskip 0.1cm}
Proton & 8.13 & 8.18\\
Neutron & 11.29 & 8.03\\
\end{tabular}
\end{ruledtabular}
\end{table}
 Figure~\ref{fig:kdist_48Ca_A_N} shows the momentum distribution of protons and neutrons in $^{48}$Ca. Momentum distributions labeled with $\mathrm{DOM_{A}}$ obtained in Ref.~\cite{calleya2025investigating} illustrate that neutrons have more high-momenta than protons, which disagrees with the results from knockout experiments and \textit{ab initio} calculations for asymmetric matter. Implementation of energy dependence in the geometry of the imaginary volume potential, both for the proton and neutron self-energy, also eliminates this discrepancy and predicts more proton high momenta than neutron high momenta as indicated by $\mathrm{DOM_{N}}$ in Fig.~\ref{fig:kdist_48Ca_A_N}. 
 We follow here the convention to consider momenta above 1.52 fm$^{-1}$ to be related to SRC~\cite{Hen:2014momentum, Duer:2018, Duer:2019direct}.
 We note that this more reasonable behavior of the neutron high momenta in ${}^{48}$Ca is accompanied by a neutron skin of 0.122 fm in excellent agreement with the experimental result of Ref.~\cite{adhikari2022precision}.
 
\section{Conclusions}
\label{sec:conclusion}

In this article, we successfully describe the high-momentum behavior captured in knockout experiments by extending the DOM to include an energy-dependent geometry. 
We accomplished this by employing an energy-dependent radius, Eq.~\eqref{eq:RE}, in the imaginary volume part of the self-energy below the Fermi energy. An energy-dependent geometry is inevitable according to previous \textit{ab initio} calculations of self-energies in finite nuclei. 

The inclusion of an energy-dependent radius breaks down the standard factorization of the potential (see Eq.~\eqref{eq:volumeS}) which complicates the evaluation of the corresponding PV integral needed to satisfy dispersion of the real part. We thus altered the implementation of the DOM to include a numerical evaluation of the PV integral, as discussed in Sec.~\ref{sec:geo_energy_dependence}. After benchmarking this new implementation, $\mathrm{DOM_N}$, with the older implementation, $\mathrm{DOM_A}$, in $^{40}\mathrm{Ca}$, we ensured that the new DOM potential still reproduces all bound and scattering observables traditionally described by the DOM in both $^{40}$Ca and $^{48}$Ca. 

The new DOM implementation causes the spectral strengths in \cref{fig:spectral_p_40Ca,fig:spectral_p_48Ca} to shift deeper in energy, bringing the deeply bound broad peak of the $0s\frac{1}{2}$ orbital closer to its corresponding experimental measurement. As expected, the description of high-momentum content is also improved in the new DOM implementation. This is evident in the full spectral function, $S(k;E)$, plotted in \cref{fig:Spectral_Strength_40Ca,fig:Spectral_Strength_48Ca} where the peaks of $S(k;E)$ increase in $E_m$ as $p_m$ increases. This behavior, which was not captured in the previous DOM iteration, is consistent with the behavior seen from knockout measurements~\cite{Rohe:2004}. 
Self-energy calculations of finite nuclei and nuclear matter considerations lead to the expectation that the peak $E_m$ will have a quadratic dependence on $p_m$.
This is confirmed by the present results although they do not match the most naive picture due to the additional complications of finite nuclei. 

The improved description of the high-momentum behavior of spectral functions naturally extends to momentum distributions. The DOM-predicted high-momentum content of protons and neutrons in $^{48}$Ca (see Fig.~\ref{fig:kdist_48Ca_A_N} and Table.~\ref{tab:high_momentum_Ca48_A_N}) is now consistent with the $np$ dominance picture established by high-momentum knockout measurements. This improvement is a natural result, since the relation between $p_m$ and $E_m$ is improved, and we demonstrated that high-momentum content is a result of the same many-body correlations that produce the spectral strength at continuum energies (large missing energies) in Fig.~\ref{fig:kdist_Ecutoff}.  

With the new DOM implementation validated in $^{40}$Ca and $^{48}$Ca and providing a consistent picture of many-body correlations, short-range correlations, and high-momentum quantities, this extension paves the way toward further studies of short-range correlations in nuclei across the nuclear chart~\cite{Nguyen2026}. This new feature, together with the growing number of nuclei that can now be described with the DOM~\cite{Ramon:2026B}, will extend its reach to better understand nuclear structure at all relevant energy scales in a consistent picture. 

\begin{acknowledgments}
This work was supported by the U.S. National Science Foundation under grants PHY-2207756 and PHY-2512895.
\end{acknowledgments}

\section*{Appendices}
\appendix 
In the following appendices, the parametrization of standard DOM, bound-state properties of $^{48}$Ca, scattering observables, and the final sets of parameters for both $^{40}$Ca and $^{48}$Ca are presented. 

\section{Parametrization of the potential}
\label{app:parametrization}

We provide a detailed description of the parametrization of the proton and neutron self-energies used in this current analysis of
$^{40}$Ca and $^{48}$Ca to fits to the bound and scattering data. The functional forms are equivalent to
those used for the $^{48}$Ca potential, detailed in Ref.~\cite{Atkinson:2019}. Parameters
which are allowed to be different for protons and neutrons will contain $(n,p)$ notation.  Asymmetry
terms have been added to the amplitudes of many of the components in the form
$\pm V_{(p,n)}\frac{N-Z}{A}$ where here only, the $+$ refers to protons and $-$ to neutrons.
Elsewhere, $\pm$ in superscripts and subscripts refer to above ($+$) and below ($-$) the Fermi
energy, $\varepsilon_F$. 
We use a simple Gaussian nonlocality in all instances \cite{Perey:1962} and restrict the nonlocal contributions to the HF term and to the volume and surface contributions to the imaginary part of the potential.
We write the HF self-energy term in the following form
with spin-orbit and a local  Coulomb contribution.
\begin{equation}
\Sigma_{HF}(\bm{r},\bm{r}') = \Sigma^{nl}_{HF}(\bm{r},\bm{r}') + V^{nl}_{so}(\bm{r},\bm{r}') + \delta(\bm{r}-\bm{r}') V_C(r),
\end{equation}
The nonlocal term is split into a volume and a narrower Gaussian term of opposite sign to make the final potential have a wine-bottle shape.
\begin{equation}
\Sigma_{HF}^{nl}\left( \bm{r},\bm{r}' \right) = -V_{HF}^{vol}\left( \bm{r},\bm{r}'\right) + V_{HF}^{wb}(\bm{r},\bm{r}'),
\end{equation}
where the volume term is given by
\begin{align}
   &V_{HF}^{vol}\left( \bm{r},\bm{r}' \right) =  V^{HF}_{sym} ,f \left (
   \tilde{r},r^{HFsym}_{(p,n)},a^{HFsym}_{(p,n)} \right ) \label{eq:HFv}\\
   &\times\left [ \frac{1}{1+x_{sym}} H \left( \bm{s};\beta^{vol_1}_{sym} \right) + \frac{x_{sym}}{1+x_{sym}} H \left( \bm{s};\beta^{vol_2}_{sym}\right) \right ]  \nonumber \\
   &\pm V_{(p,n)}^{HFasy} \frac{N-Z}{A}f \left ( \tilde{r},r^{HFaym}_{(p,n)},a^{HFasy}_{(p,n)} \right
   )\times \nonumber \\
   &\left [ \frac{1}{1+x_{sym}} H \left( \bm{s};\beta^{vol_1asy}_{(p,n)} \right)  +
   \frac{x_{sym}}{1+x_{sym}} H \left( \bm{s};\beta^{vol_2asy}_{sym}\right) \right ].\nonumber
\end{align} 
allowing for two different nonlocalities with different weights ($0 \le x_{sym} \le1$ in
Eq.~\eqref{eq:HFv}). 
With   the notation $\tilde{r} =(r+r')/2$ and $\bm{s}=\bm{r}-\bm{r}'$,
the wine-bottle ($wb$) shape is described by
\begin{equation}
V_{HF}^{wb}(\bm{r},\bm{r}') = V^{wb}_{(p,n)}  \exp{\left(- \tilde{r}^2/(\rho^{wb}_{(p,n)})^2\right)} H \left( \bm{s};\beta^{wb} \right ),
\label{eq:wb}
\end{equation}
where the nonlocality in Eq.~\eqref{eq:wb} is represented by a Gaussian form
\begin{equation}
H \left( \bm{s}; \beta \right) = \exp \left( - \bm{s}^2 / \beta^2 \right)/ (\pi^{3/2} \beta^3) .
\nonumber
\end{equation}
As usual, we employ Woods-Saxon form factors 
\begin{align}
f(r,r_{i},a_{i})=\left[1+\exp \left({\frac{r-r_{i}A^{1/3}}{a_{i}}%
}\right)\right]^{-1} .
\label{Eq:WS}
\end{align}
The Coulomb term is obtained from the experimental charge density distribution for $^{48}$Ca~\cite{de1987nuclear}.

The local spin-orbit interaction is given by
\begin{align}
      V_{so}(\bm{r},\bm{r'})= \left( \frac{\hbar}{m_{\pi }c}\right) ^{2}
      V^{so}_{(p,n)}\frac{1}{\tilde{r}}\frac{d}{d\tilde{r}}f(\tilde{r},r^{so}_{(p,n)},a^{so})\nonumber \\
      \times\bm{\ell}\cdot \bm{\sigma}
      H(\bm{s};\beta^{so}), 
   \label{eq:HFso}
\end{align}
where $\left( \hbar /m_{\pi }c\right) ^{2}$=2.0~fm$^{2}$ 
as in Ref.~\cite{Mueller:2011}.

The fully-nonlocal imaginary part of the DOM self-energy has the following form,
\begin{align}
\label{eq:imnl}
&\textrm{Im}\ \Sigma^{nl}(\bm{r},\bm{r}';E) = \hspace{5cm} \\ \nonumber  
&-W^{vol}_{0\pm}(E) f\left(\tilde{r};r^{vol}_{\pm (p,n)};a^{vol}_{\pm}\right)H \left( \bm{s}; \beta_{\pm (p,n)}^{vol}\right) \hspace{1cm} \\ \nonumber
&+ 4a^{sur}_{\pm sym}W^{sur0}_{\pm}\left( E\right)H \left( \bm{s}; \beta_{\pm}^{sur0}\right) \frac{d}{d \tilde{r} }f(\tilde{r},r^{sur0}_{\pm(p,n))},a^{sur}_{\pm sym}) \\ \nonumber
& + 4 a^{sur}_{\pm sym} W^{sur}_{\pm}(E) H \left( \bm{s};\beta^{sur}_{\pm (p,n)} \right ) \frac{d}{d \tilde{r} }f(\tilde{r},r^{sur}_{\pm(p,n)},a^{sur}_{\pm(p,n)}) \\ \nonumber 
&+ \textrm{Im}\Sigma_{so}(\bm{r},\bm{r}';E).
\end{align}
Note that the parameters relating to the shape of the imaginary spin-orbit term
are the same as those used for the real spin-orbit term.
At energies well removed
from $\varepsilon_F$, the form of the imaginary volume potential should not be
symmetric about $\varepsilon_F$ as indicated by the $\pm$ notation in the subscripts and superscripts~\cite{Dussan:2011}.
While more symmetric about $\varepsilon_F$, we have allowed a similar option for the surface absorption that is also supported by theoretical work reported in Ref.~\cite{Waldecker:2011}.
Allowing for the aforementioned asymmetry around $\varepsilon_F$ the following form was assumed for the depth of the volume potential~\cite{Mueller:2011}
\begin{widetext} 
\begin{equation}
W^{vol}_{0\pm}(E) =  \Delta W^{\pm}_{NM}(E) +  
\begin{cases}
0 & \text{if } |E-\varepsilon_F| < \mathcal{E}^{vol}_{\pm} \\
A^{vol}_{\pm(p,n)}  \frac{\left(|E-\varepsilon_F|-\mathcal{E}^{vol}_{\pm}\right)^4}
{\left(|E-\varepsilon_F|-\mathcal{E}^{vol}_{\pm}\right)^4 + (B^{vol}_{\pm})^4} & 
 \text{if } |E-\varepsilon_F| > \mathcal{E}^{vol}_{\pm} ,
\end{cases} 
\label{eq:volumeS}
\end{equation}
\end{widetext}
and $\Delta W^{\pm}_{NM}(E)$ in Eq.~\eqref{eq:volumeS} is the energy-asymmetric correction modeled after
nuclear-matter calculations. The asymmetry above and below $\varepsilon_F$ is essential to accommodate the Jefferson Lab $(e,e'p)$ data at large missing energy.
The energy-asymmetric correction was taken as 
\begin{widetext} 
\begin{equation}
\Delta W^{\pm}_{NM}(E)=
\begin{cases}
\alpha^{sym}_{(p,n)} A^{vol}_{+(p,n)}\left[ \sqrt{E}+\frac{\left( \varepsilon_F+\mathbb{E}_{+}\right) ^{3/2}}{2E}-\frac{3}{2}
\sqrt{\varepsilon_F+\mathbb{E}_{+}}\right] & \text{for }E-\varepsilon_F>\mathbb{E}_{+} \\ 
-  A^{vol}_{-(p,n)} \frac{(\varepsilon_F-E-\mathbb{E}_{-})^2}{(\varepsilon_F-E-\mathbb{E}_{-})^2+(\mathbb{E}_{-})^2} & \text{for }E-\varepsilon_{F}<-\mathbb{E}_{-} \\ 
0 & \text{otherwise},
\end{cases} 
\label{eq:Wnmnl}
\end{equation} 
\end{widetext}
where $E$ in Eq.~\eqref{eq:Wnmnl} corresponds to the center-of-mass energy.

To describe the energy dependence of surface absorption, we employed the form of
Ref.~\cite{Charity:2007}, but include two components, one with symmetric parameters, the other with
asymmetric parameters.
\begin{align}
W^{sur0}_{\pm}\left( E\right) =\omega _{4}(E,A^{sur0}_{\pm},B^{sur0_1}_{\pm},0)- \nonumber \\
\omega_{2}(E,A^{sur0}_{\pm},B_{\pm}^{sur0_2},C^{sur0}_{\pm}),  \label{eq:paranl} 
\end{align}
\begin{align}
  W^{sur}_{\pm(p,n))}\left( E\right) =\omega _{4}(E,A^{sur}_{\pm(p,n)},B^{sur_1}_{\pm(p,n)},0)- \nonumber \\
  \omega_{2}(E,A^{sur}_{\pm(p,n)},B_{\pm(p,n)}^{sur_2},C^{sur}_{\pm(p,n)}),  \label{eq:paran2} 
\end{align}
where the $\omega$ functions in Eqs.~\eqref{eq:paranl} and~\eqref{eq:paran2} are defined as
\begin{align}
\omega _{n}(E,A^{sur},B^{sur},C^{sur})=A^{sur}\;\Theta \left(
X\right) \frac{X^{n}}{X^{n}+\left( B^{sur}\right) ^{n}}, \nonumber \\
\label{eq:omega}
\end{align}%
and $\Theta \left( X\right) $ is Heaviside's step function and $%
X=\left\vert E-\varepsilon_F\right\vert -C^{sur}$. 

The imaginary spin-orbit term in Eq.~\eqref{eq:imnl} has the same form as the real spin-orbit
term in Eq.~\eqref{eq:HFso},
\begin{align}
      W_{so}(\bm{r},\bm{r'};E)= \left( \frac{\hbar}{m_{\pi }c}\right) ^{2}
      W^{so}(E)\frac{1}{\tilde{r}}\frac{d}{d\tilde{r}}f(\tilde{r},r^{so}_{(p,n)},a^{so})\nonumber \\
      \times\bm{\ell}\cdot \bm{\sigma}
      H(\bm{s};\beta^{so}), 
   \label{eq:imag_so}
\end{align}
where the radial parameters for the imaginary component are the same as those used for the real part of the spin-orbit potential. 
It is important to note that $\textrm{Im}\Sigma_{so}$ grows with increasing $\ell$, and for large $\ell$ this can lead to an inversion of the sign of the self-energy, which results in negative occupation. While the
form of Eq.~\eqref{eq:HFso} suppresses this behavior, it is still not a proper solution. One must be careful that the magnitude of $W_{so}(E)$ does not exceed that of the volume and surface
components. As the imaginary spin-orbit component is
generally needed only at high energies, the form of Ref.~\cite{Mueller:2011} is employed, 
\begin{equation}
   W^{so}(E)= A^{so}  \frac{(E-\varepsilon_F)^4}{(E-\varepsilon_F)^4+(B^{so})^4} .
   \label{eq:ImSO}
\end{equation}%

With Eq.~\eqref{eq:ImSO}, all ingredients of the self-energy have now been identified and their functional form described.
In addition to the Hartree-Fock contribution and the absorptive potentials we also include the
dispersive real part from all imaginary contributions according to Eq.(1).
\section{Bound-State Observables}
\label{app:bound_state}

\subsection{\texorpdfstring{$^{40}$Ca observables}{}}
\label{app:bound_state40Ca}
$^{40}$Ca bound-state observables are presented in Sec.III~B of the main article to show that the new implementation in dispersive optical model $(\mathrm{DOM_N})$ predicts experimental data with the same quality as calculated using standard DOM implementation $(\mathrm{DOM_A})$.

\subsection{\texorpdfstring{$^{48}$Ca observables}{}}
\label{app:bound_state48Ca}
\begin{figure}[b]
    \includegraphics[width=\columnwidth]{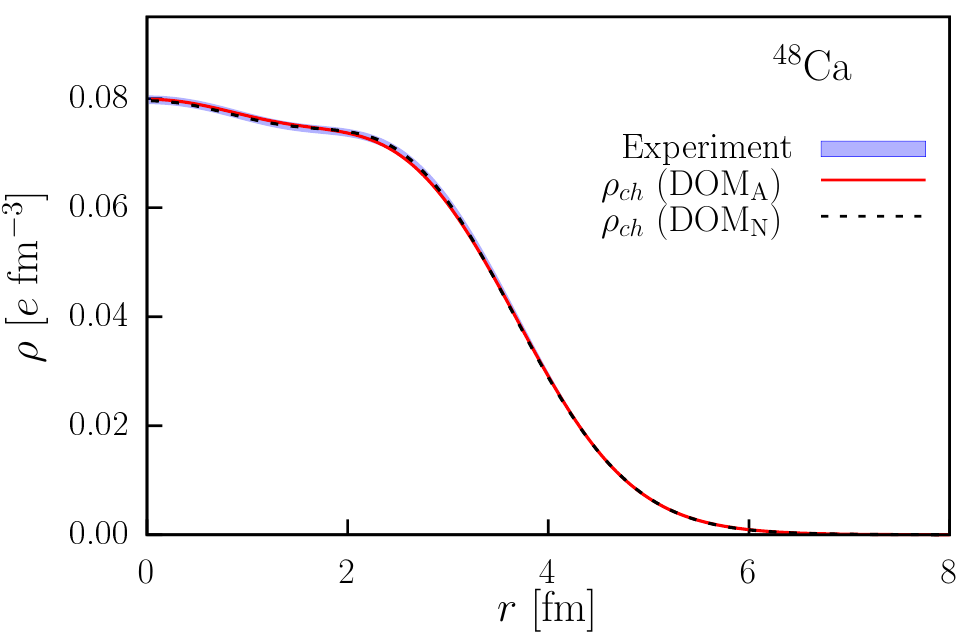}
    \caption{Experimental charge density of $^{48}$Ca~\cite{deVries:1987} compared with the DOM-calculated charge density.$\mathrm{DOM_A}$ represents the charge density calculated when the geometry of the volume imaginary has no energy dependence. $\mathrm{DOM_N}$  shows a similar quality of charge density distribution with an energy-dependent geometry of the volume imaginary part of the self-energy.}
   \label{fig:chd}
\end{figure}
The energy-dependent geometry is introduced only for the volume-imaginary part of the self-energy in the negative energy domain. Hence, it is important to verify that this extension to the DOM preserves the ability to describe ground-state experimental observables.
For ${}^{48}$Ca, the $\mathrm{DOM_{A}}$ and $\mathrm{DOM_{N}}$ calculated total binding energy are $ E_0^\mathrm{DOM_{A}} = 429.25$~MeV and  $E_0^\mathrm{DOM_{N}} = 420.64$~MeV, respectively, which are both close to the experimental value of $E_0^\mathrm{Exp} = 415.968$~MeV~\cite{wang2021ame}. Similarly, the RMS charge radii also have values $r^\mathrm{RMS}_\mathrm{DOM_{A}} = 3.485$~fm and $r^\mathrm{RMS}_\mathrm{DOM_{N}} = 3.479$~fm, compared to the experimental value of $r^\mathrm{RMS}_\mathrm{Exp} = 3.4771$~fm~\cite{Angeli:2013}.
\begin{table}[ht]
\centering
\caption{Comparison of proton and neutron single-particle levels (in MeV) in $^{48}$Ca. Here $\mathrm{DOM_A}$ represents the levels calculated using the analytical version of DOM, and $\mathrm{DOM_N}$ indicates levels calculated with energy-dependent geometry of volume imaginary.}
\label{tab:levels}
\begin{ruledtabular}
\renewcommand{\arraystretch}{1.3}
\begin{tabular}{l|cccccc}

& \multicolumn{3}{c}{Proton levels [MeV]} & \multicolumn{3}{c}{Neutron levels [MeV]} \\

\cline{2-4} \cline{5-7}

$n\ell j$ & Exp. & $\mathrm{DOM_A}$ & $\mathrm{DOM_N}$ & Exp. & $\mathrm{DOM_A}$ & $\mathrm{DOM_N}$ \\

\cline{1-4} \cline{5-7}
$0d\frac{5}{2}$ & $-19.20$ & $-22.82$  & $-22.18$   & $-$      & $-$          & $-$ \\
$0d\frac{3}{2}$ & $-16.10$ & $-16.86$  & $-16.12$   & $-12.54$ & $-13.55$     & $-13.84 $ \\
$1s\frac{1}{2}$ & $-15.80$ & $-16.79$  & $-16.02$   & $-12.54$ & $-17.24$     & $17.67$ \\
$0f\frac{7}{2}$ & $-9.62$  & $-10.69$  & $-9.98$    & $-9.95$  & $-10.59$     & $-10.75$\\
$1p\frac{3}{2}$ & $-6.54$  & $-5.22$   & $-4.07$    & $-5.14$  & $-5.84$      & $-5.80 $ \\
$0f\frac{7}{2}$ & $-4.92$  & $-4.35$   & $-3.25$    & $-1.19$  & $-1.30$      & $-1.23$\\
$1p\frac{1}{2}$ & $-3.58$  & $-2.94$   & $-1.81$    & $-3.11$  & $-3.17$      &$-3.07 $ \\
\end{tabular}
\end{ruledtabular}
\end{table}
\begin{figure}[b]
    \includegraphics[width=\columnwidth]{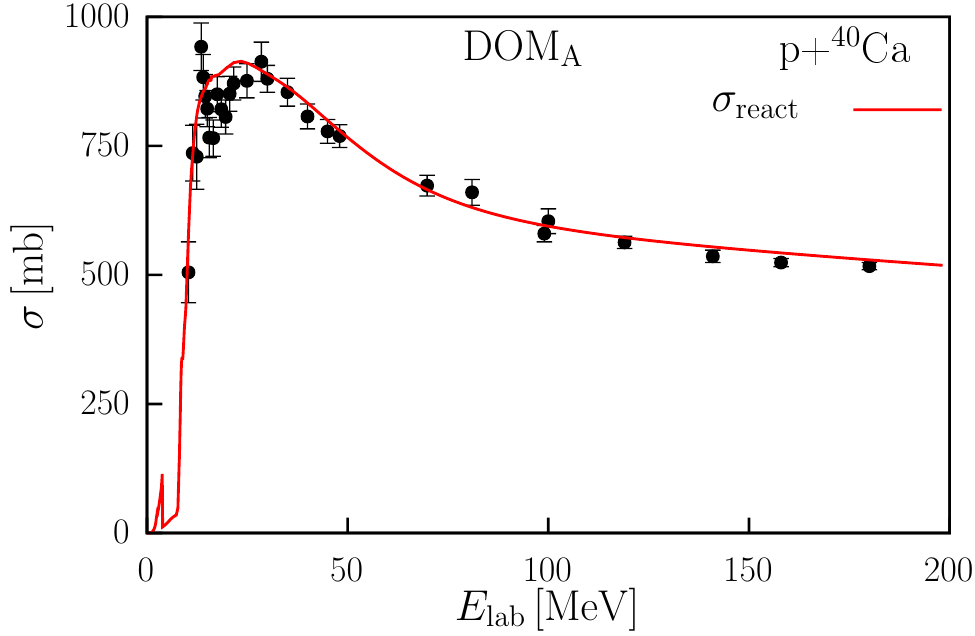}
      \caption{$\mathrm{DOM_{A}}$ calculated proton reaction cross section for $^{40}$Ca.}
   \label{fig:react_p_DOM_A}
\end{figure}
\begin{figure}[b]
    \includegraphics[width=\columnwidth]{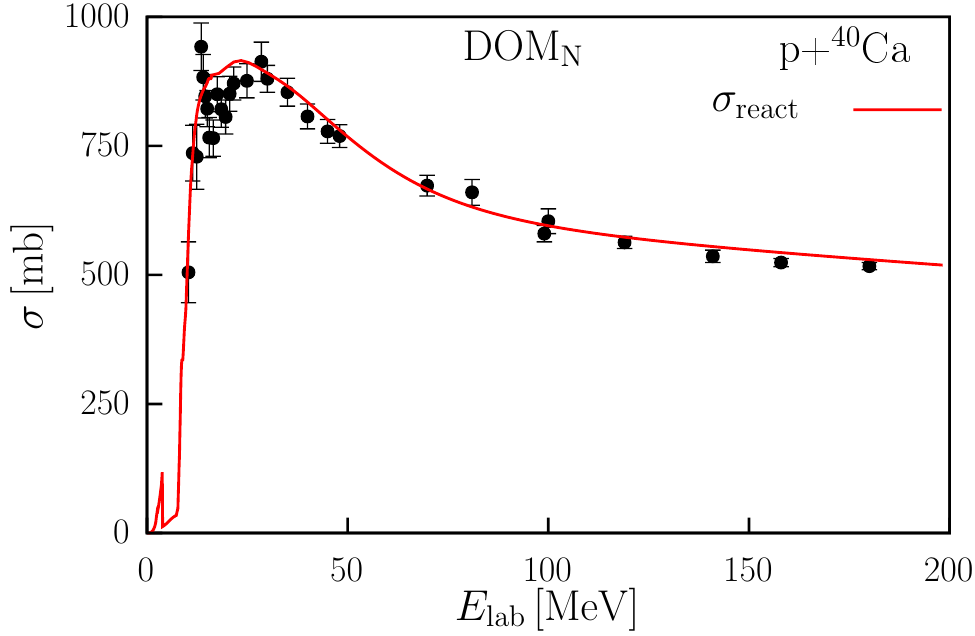}
      \caption{$\mathrm{DOM_{N}}$ calculated proton reaction cross section for $^{40}$Ca.}
   \label{fig:react_p_DOM_N}
\end{figure}
\section{Scattering observables}
\label{app:scattering}
The dispersive optical model connects the energy domain below the Fermi energy to the domain above through the subtracted dispersion relation in Eq.~\eqref{eq:dispersion}. In this article, the energy dependence in the geometry of the volume imaginary part below the Fermi energy of the self-energy is implemented. Consequently, the self-energy, as well as the bound-state properties, are expected to be affected more than the scattering observables, which lie in the energy domain above the Fermi energy. The effect of the $\mathrm{DOM_N}$ implementation on the bound-state properties was discussed in the main article in Sec.~III~B, and plots associated with the scattering observables are presented. All the experimental data are from Ref.\cite{Mueller:2011} except for the charge density.

We offset the differential cross section data at each energy by factors of 10 to visualize all data (filled circles) and results (solid lines) simultaneously in Figs.Figs.~\ref{fig:xsec_p_n_DOM_A}, \ref{fig:xsec_p_n_DOM_N}, \ref{fig:xsec_p_n_DOM_A_48Ca}, and \ref{fig:xsec_p_n_DOM_N_48Ca}.

In this appendix, figures labeled with $\mathrm{DOM_A}$ represent the fitted results from the standard DOM, and $\mathrm{DOM_N}$ represents refitted results after the implementation of the energy dependence in the geometry of the volume imaginary part of the self-energy.

\subsection{${}^{40}$Ca observables}
\begin{figure}[t]
    \includegraphics[width=\columnwidth]{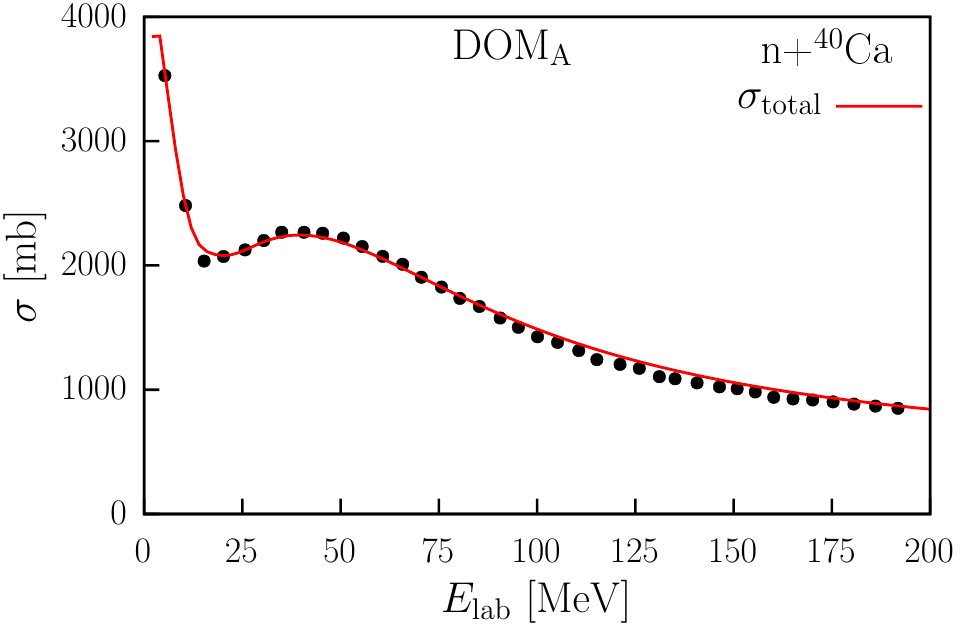}
      \caption{$\mathrm{DOM_{A}}$ calculated total cross section for neutrons scattered off $^{40}$Ca.}
   \label{fig:total_n_DOM_A}
\end{figure}

\begin{figure}[t]
      \includegraphics[width=\columnwidth]{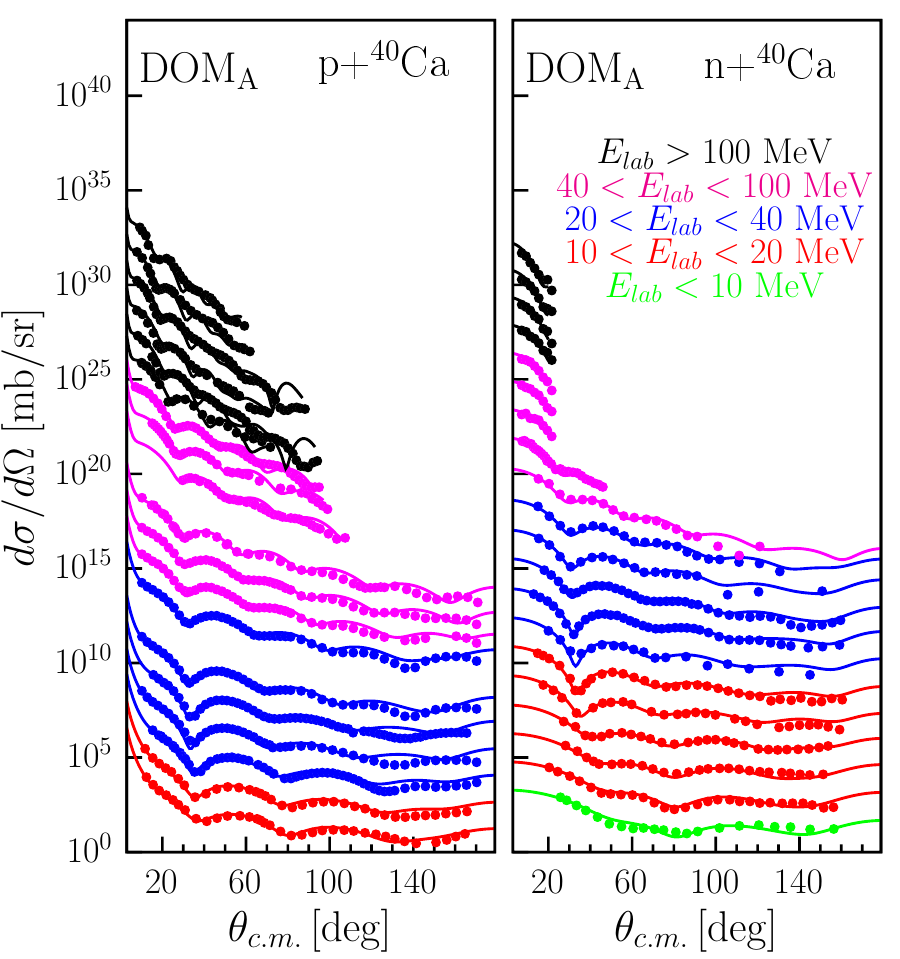}
      \captionof{figure}{$\mathrm{DOM_A}$ calculated angular distributions the differential cross section of $p$/$n$ elastic scattering off $^{40}$Ca.}
   \label{fig:xsec_p_n_DOM_A}
\end{figure}

\begin{figure}[t]
    \includegraphics[width=\columnwidth]{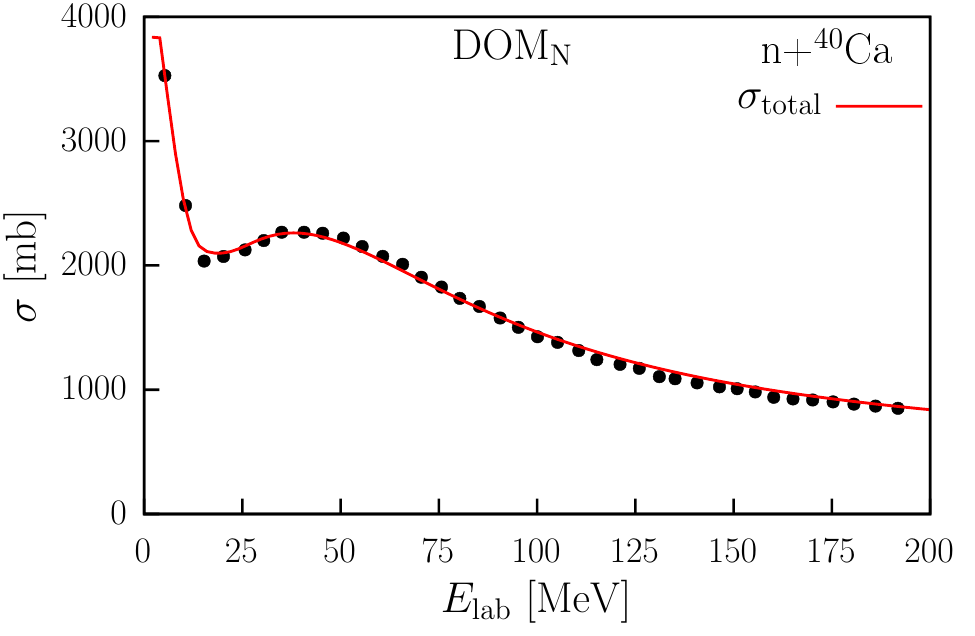}
      \caption{$\mathrm{DOM_{N}}$ calculated total cross section for neutrons scattered off $^{40}$Ca.}
   \label{fig:total_n_DOM_N}
\end{figure}

\begin{figure}[t]
\includegraphics[width=\columnwidth]{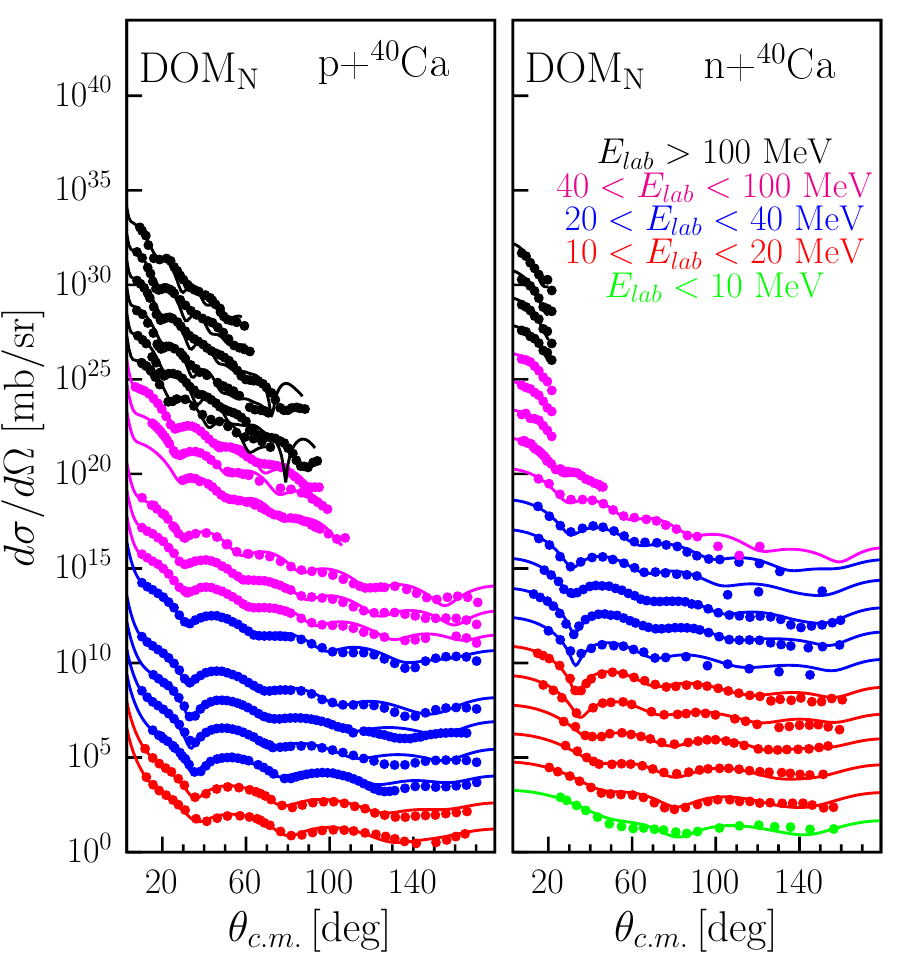}
\captionof{figure}{$\mathrm{DOM_N}$ calculated angular distributions the differential cross section of $p$/$n$ elastic scattering off $^{40}$Ca.}
\label{fig:xsec_p_n_DOM_N}
\end{figure}

Fig.\ref{fig:react_p_DOM_A} and Fig.\ref{fig:react_p_DOM_N} represents the proton reaction cross section for $^{40}$Ca. Total cross section for neutrons scattered off $^{40}$Ca is depicted in Fig.\ref{fig:total_n_DOM_A} and Fig.\ref{fig:total_n_DOM_N} respectively. Fig.\ref{fig:xsec_p_n_DOM_A} and Fig.\ref{fig:xsec_p_n_DOM_N} show the angular distribution of the differential cross section for both proton and neutron. All the observables have plots calculated with $\mathrm{DOM_A}$ and $\mathrm{DOM_N}$ to emphasize the reproducibility of the same quality of fit.
\subsection{${}^{48}$Ca observables}
\begin{figure*}[p]
\centering
\begin{minipage}[t]{0.48\textwidth}
    \centering
    \includegraphics[width=\textwidth]{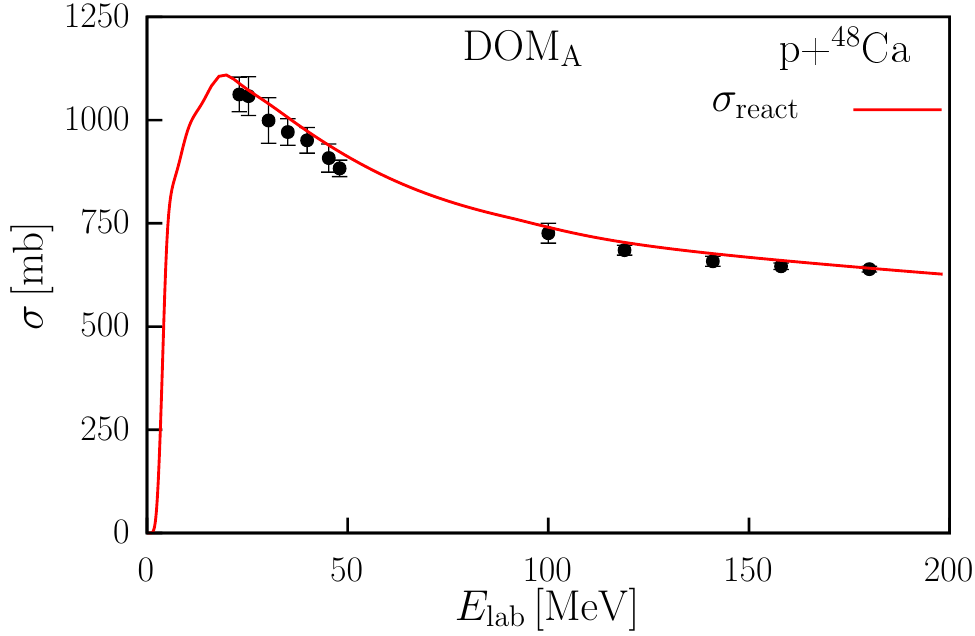}
    \captionof{figure}{$\mathrm{DOM_{A}}$ calculated proton reaction cross section for $^{48}$Ca.}
   \label{fig:react_p_DOM_A_48Ca}
\vspace{1em}
      \includegraphics[width=\textwidth]{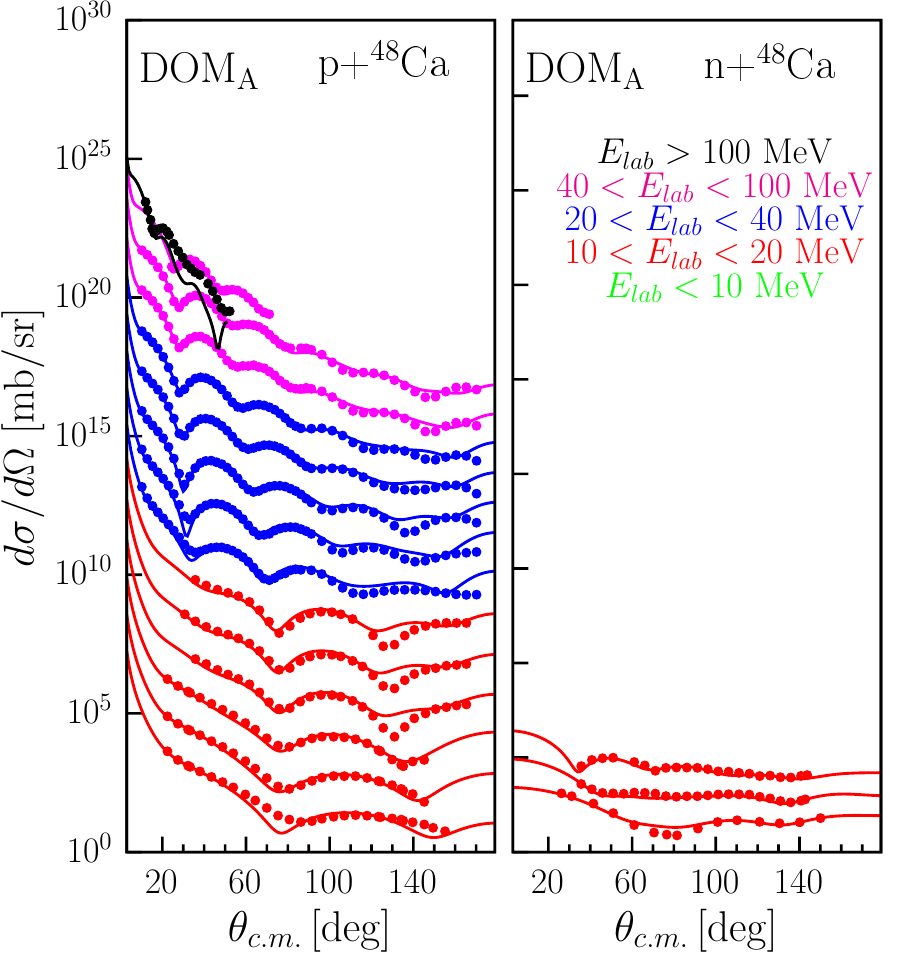}
      \captionof{figure}{$\mathrm{DOM_A}$ calculated angular distributions the differential cross section of $p$/$n$ elastic scattering off $^{48}$Ca.}
   \label{fig:xsec_p_n_DOM_A_48Ca}
\vspace{1em}
    \includegraphics[width=\textwidth]{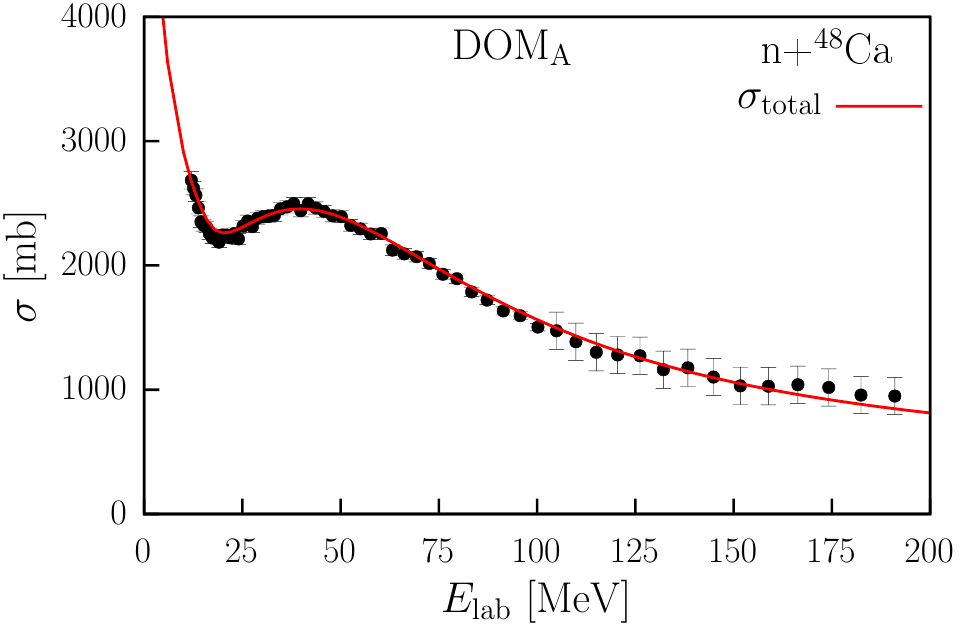}
      \captionof{figure}{$\mathrm{DOM_{A}}$ calculated total cross section for neutrons scattered off $^{48}$Ca.}
   \label{fig:total_n_DOM_A_48Ca}
\end{minipage}
\hfill
\begin{minipage}[t]{0.48\textwidth}
\centering
    \includegraphics[width=\textwidth]{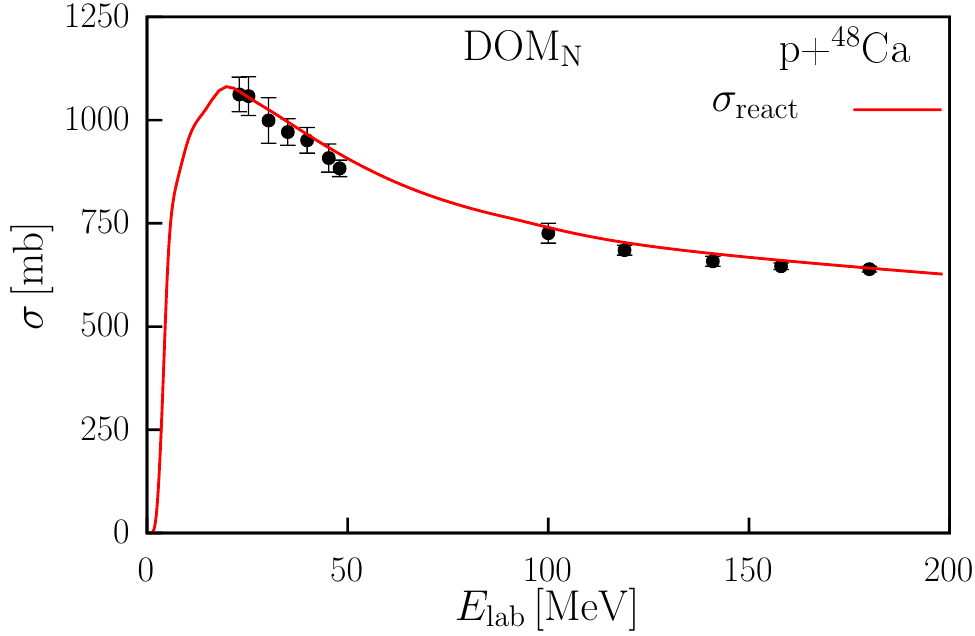}
      \captionof{figure}{$\mathrm{DOM_{N}}$ calculated proton reaction cross section for $^{48}$Ca.}
   \label{fig:react_p_DOM_N_48Ca}
      \includegraphics[width=\textwidth]{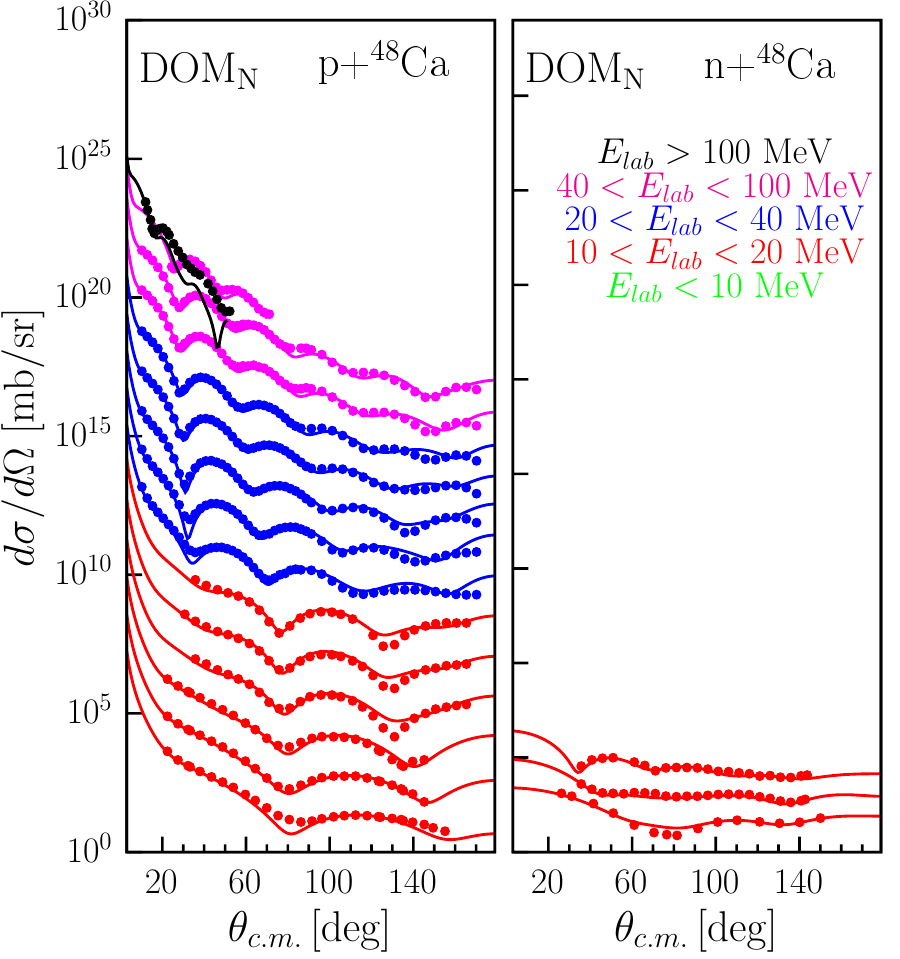}
      \captionof{figure}{$\mathrm{DOM_N}$ calculated angular distributions the differential cross section of $p$/$n$ elastic scattering off $^{48}$Ca.}
   \label{fig:xsec_p_n_DOM_N_48Ca}
\vspace{1em}
    \includegraphics[width=\textwidth]{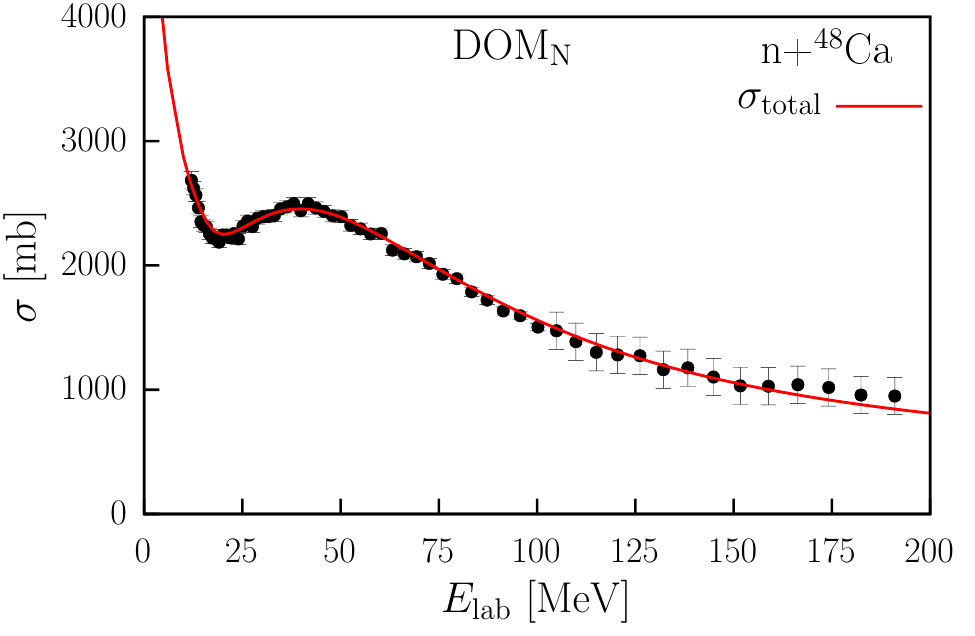}
      \captionof{figure}{$\mathrm{DOM_{N}}$ calculated total cross section for neutrons scattered off $^{48}$Ca.}
   \label{fig:total_n_DOM_N_48Ca}
\end{minipage}

\end{figure*}

Fig.\ref{fig:react_p_DOM_A_48Ca} and Fig.\ref{fig:react_p_DOM_N_48Ca} represents the proton reaction cross section for $^{48}$Ca. Fig.\ref{fig:xsec_p_n_DOM_A_48Ca} and Fig.\ref{fig:xsec_p_n_DOM_N_48Ca} show the angular distribution of the differential cross section for both proton and neutron. Total cross section for neutrons scattered off $^{48}$Ca is depicted in Fig.\ref{fig:total_n_DOM_A_48Ca} and Fig.\ref{fig:total_n_DOM_N_48Ca} respectively. All the observables have plots calculated with $\mathrm{DOM_A}$ and $\mathrm{DOM_N}$ to emphasize the reproducibility of the same quality of fit.

\section{Parameters}
\label{app:params}

The parameters used for the symmetric part of the self-energy are presented in
Table~\ref{Tbl:symmetric_40Ca} for $^{40}$Ca and in
Table~\ref{Tbl:symmetric_48Ca} for $^{48}$Ca. All asymmetric parameters are presented in
Table~\ref{Tbl:asymmetric_40Ca} for $^{40}$Ca and in
Table~\ref{Tbl:asymmetric_48Ca} for $^{48}$Ca . There are 30 Lagrange-Legendre and Lagrange-Laguerre grid points used in the calculations~\cite{Baye_review,Baye:2010}. For $^{40}$Ca, the scaling parameter for the Lagrange-Laguerre mesh points is $a_L=0.15$. The matching radius used for scattering calculations is $a=12$~fm. Parameters shown in parentheses indicate that they were reset to the corresponding non-parenthesized parameter during the refitting performed with $\mathrm{DOM_N}$. Parameters without a parenthesized counterpart were used identically in both $\mathrm{DOM_A}$ and $\mathrm {DOM_N}$.
\begin{table}[H]
\caption{Fitted parameter values for proton and neutron potentials for $^{40}$Ca. Asterisked $(*)$ parameters are set to zero, causing the associated terms to vanish because $^{40}$Ca is a symmetric nucleus. Related parameters set to 1.000 is inconsequential, as the whole term is zero because of the asterisked $(*)$ parameters.}
\label{Tbl:asymmetric_40Ca}%
\begin{ruledtabular}
\begin{tabular}{ccSSc} 
Parameter & Unit &  {\text{$p$ value}}    &  {\text{$n$ value}} & Eq. No. \\

\hline
\multicolumn{5}{c}{Hartree-Fock} \\
\hline
$r^{HFsym}_{(p,n)}$ & [fm] & 1.149 & 1.149 & (\ref{eq:HFv}) \\
$(r^{HFsym}_{(p,n)})$ & [fm] & 1.155 & 1.155 & (\ref{eq:HFv}) \\
$a^{HFsym}_{(p,n)}$ & [fm] & 0.643 & 0.643 & (\ref{eq:HFv}) \\
$(a^{HFsym}_{(p,n)})$ & [fm] & 0.633 & 0.633 & (\ref{eq:HFv}) \\
$V_{(p,n)}^{wb}$ & [MeV] & 30.038 & 30.038 & (\ref{eq:wb}) \\
$\rho_{(p,n)}^{wb}$ & [MeV] & 0.517 & 0.517 & (\ref{eq:wb}) \\
$^{*}V^{HFasy}_{(p,n)}$ & [MeV] & 0.000 & 0.000 & (\ref{eq:HFv}) \\
$r^{HFasy}_{(p,n)}$ & [fm] & 1.000 & 1.000 & (\ref{eq:HFv}) \\
$a^{HFasy}_{(p,n)}$ & [fm] & 1.000 & 1.000 & (\ref{eq:HFv}) \\
$\beta^{vol_1asy}_{(p,n)}$ & [fm] & 1.000 & 1.000 & (\ref{eq:HFv}) \\
$\beta^{vol_2asy}_{(p,n)}$ & [fm] & 1.000 & 1.000 & (\ref{eq:HFv}) \\
\hline
\multicolumn{5}{c}{Spin-orbit} \\
\hline
$V^{so}_{(p,n)}$ & [MeV] & 12.902 & 12.902 & (\ref{eq:HFso}) \\
$r^{so}_{(p,n)}$ & [fm] & 0.807 & 0.807 & (\ref{eq:HFso}) \\
\hline
\multicolumn{5}{c}{Volume imaginary} \\
\hline
$\beta^{vol}_{+(p,n)}$ & [fm] & 0.405 & 0.405 & (\ref{eq:imnl}) \\
$\beta^{vol}_{-(p,n)}$ & [fm] & 1.119 & 1.119 & (\ref{eq:imnl}) \\
$r^{vol}_{+(p,n)}$ & [fm] & 1.502 & 1.502 & (\ref{eq:imnl}) \\
$r^{vol}_{-(p,n)}$ & [fm] & 1.201 & 1.201 & (\ref{eq:imnl}) \\
$(r^{vol}_{-(p,n)})$ & [fm] & 1.451 & 1.451 & (\ref{eq:imnl}) \\
$A^{vol}_{+(p,n)}$ & [MeV] & 6.489 & 6.489 & (\ref{eq:volumeS}) \\
$A^{vol}_{-(p,n)}$ & [MeV] & 14.043 & 14.043 & (\ref{eq:volumeS}) \\
$(A^{vol}_{-(p,n)})$ & [MeV] & 0.943 & 0.943 & (\ref{eq:volumeS}) \\
$\alpha^{sym}_{(p,n)}$ & [fm] & 0.130 & 0.130 & (\ref{eq:Wnmnl}) \\
\hline
\multicolumn{5}{c}{Surface imaginary} \\
\hline
$\beta^{sur}_{+(p,n)}$ & [fm] & 1.000 & 1.000 & (\ref{eq:imnl}) \\
$\beta^{sur}_{-(p,n)}$ & [fm] & 1.000 & 1.000 & (\ref{eq:imnl}) \\
$r^{sur0}_{+(p,n)}$ & [fm] & 1.109 & 1.109 & (\ref{eq:imnl}) \\
$r^{sur0}_{-(p,n)}$ & [fm] & 0.590 & 0.590 & (\ref{eq:imnl}) \\
$r^{sur}_{+(p,n)}$ & [fm] & 1.000 & 1.000 & (\ref{eq:imnl}) \\
$r^{sur}_{-(p,n)}$ & [fm] & 1.000 & 1.000 & (\ref{eq:imnl}) \\
$a^{sur}_{+(p,n)}$ & [fm] & 1.000 & 1.000 & (\ref{eq:imnl}) \\
$a^{sur}_{-(p,n)}$ & [fm] & 1.000 & 1.000 & (\ref{eq:imnl}) \\
$^{*}A^{sur}_{+(p,n)}$ & [MeV] & 0.000 & 0.000 & (\ref{eq:paran2}) \\
$^{*}A^{sur}_{-(p,n)}$ & [MeV] & 0.000 & 0.000 & (\ref{eq:paran2}) \\
$B^{sur_1}_{+(p,n)}$ & [MeV] & 1.000 & 1.000 & (\ref{eq:paran2}) \\
$B^{sur_1}_{-(p,n)}$ & [MeV] & 1.000 & 1.000 & (\ref{eq:paran2}) \\
$B^{sur_2}_{+(p,n)}$ & [MeV] & 1.000 & 1.000 & (\ref{eq:paran2}) \\
$B^{sur_2}_{-(p,n)}$ & [MeV] & 1.000 & 1.000 & (\ref{eq:paran2}) \\
$C^{sur}_{+(p,n)}$ & [MeV] & 1.000 & 1.000 & (\ref{eq:paran2}) \\
$C^{sur}_{-(p,n)}$ & [MeV] & 1.000 & 1.000 & (\ref{eq:paran2}) \\
\end{tabular}
\end{ruledtabular}
\end{table}

\begin{table}[H]
\caption{Fitted parameter values for proton and neutron potentials in 
$^{48}$Ca. This table also lists the number of the equation that defines each individual parameter.}
\label{Tbl:asymmetric_48Ca}%
\begin{ruledtabular}
\begin{tabular}{ccSSc} 
Parameter & Unit &  {\text{$p$ value}}    &  {\text{$n$ value}} & Eq. No. \\

\hline
\multicolumn{5}{c}{Hartree-Fock} \\
\hline
$r^{HFsym}_{(p,n)}$ & [fm] & 1.162 & 1.095 & (\ref{eq:HFv}) \\
$(r^{HFsym}_{(p,n)})$ & [fm] & 1.150 & 1.093 & (\ref{eq:HFv}) \\
$a^{HFsym}_{(p,n)}$ & [fm] & 0.631 & 0.657 & (\ref{eq:HFv}) \\
$(a^{HFsym}_{(p,n)})$ & [fm] & 0.601 & 0.657 & (\ref{eq:HFv}) \\
$V_{(p,n)}^{wb}$ & [MeV] & 56.000 & 40.038 & (\ref{eq:wb}) \\
$(V_{(p,n)}^{wb})$ & [MeV] & 56.000 & 25.038 & (\ref{eq:wb}) \\
$\rho_{(p,n)}^{wb}$ & [MeV] & 0.727 & 0.717 & (\ref{eq:wb}) \\
$V^{HFasy}_{(p,n)}$ & [MeV] & 5.603 & 17.262 & (\ref{eq:HFv}) \\
$(V^{HFasy}_{(p,n)})$ & [MeV] & 4.003 & 13.262 & (\ref{eq:HFv}) \\
$r^{HFasy}_{(p,n)}$ & [fm] & 1.104 & 1.331 & (\ref{eq:HFv}) \\
$(r^{HFasy}_{(p,n)})$ & [fm] & 1.004 & 1.331 & (\ref{eq:HFv}) \\
$a^{HFasy}_{(p,n)}$ & [fm] & 0.500 & 0.500 & (\ref{eq:HFv}) \\
$\beta^{vol_1asy}_{(p,n)}$ & [fm] & 1.537 & 1.000 & (\ref{eq:HFv}) \\
$\beta^{vol_2asy}_{(p,n)}$ & [fm] & 1.000 & 1.000 & (\ref{eq:HFv}) \\
\hline
\multicolumn{5}{c}{Spin-orbit} \\
\hline
$V^{so}_{(p,n)}$ & [MeV] & 12.902 & 17.972 & (\ref{eq:HFso}) \\
$r^{so}_{(p,n)}$ & [fm] & 0.807 & 0.873 & (\ref{eq:HFso}) \\
\hline
\multicolumn{5}{c}{Volume imaginary} \\
\hline
$\beta^{vol}_{+(p,n)}$ & [fm] & 0.405 & 0.405 & (\ref{eq:imnl}) \\
$\beta^{vol}_{-(p,n)}$ & [fm] & 1.119 & 1.119 & (\ref{eq:imnl}) \\
$r^{vol}_{+(p,n)}$ & [fm] & 1.463 & 1.591 & (\ref{eq:imnl}) \\
$r^{vol}_{-(p,n)}$ & [fm] & 1.101 & 1.136 & (\ref{eq:imnl}) \\
$(r^{vol}_{-(p,n)})$ & [fm] & 1.341 & 1.336 & (\ref{eq:imnl}) \\
$A^{vol}_{+(p,n)}$ & [MeV] & 4.726 & 3.488 & (\ref{eq:volumeS}) \\
$A^{vol}_{-(p,n)}$ & [MeV] & 13.743 & 10.528 & (\ref{eq:volumeS}) \\
$(A^{vol}_{-(p,n)})$ & [MeV] & 00.743 & 00.528 & (\ref{eq:volumeS}) \\
$\alpha^{sym}_{(p,n)}$ & [fm] & 0.161 & 0.135 & (\ref{eq:Wnmnl}) \\
\hline
\multicolumn{5}{c}{Surface imaginary} \\
\hline
$\beta^{sur}_{+(p,n)}$ & [fm] & 1.377 & 1.000 & (\ref{eq:imnl}) \\
$\beta^{sur}_{-(p,n)}$ & [fm] & 1.200 & 1.400 & (\ref{eq:imnl}) \\
$r^{sur0}_{+(p,n)}$ & [fm] & 1.054 & 1.116 & (\ref{eq:imnl}) \\
$r^{sur0}_{-(p,n)}$ & [fm] & 0.590 & 0.655 & (\ref{eq:imnl}) \\
$(r^{sur0}_{-(p,n)})$ & [fm] & 0.700 & 0.655 & (\ref{eq:imnl}) \\
$r^{sur}_{+(p,n)}$ & [fm] & 1.000 & 1.100 & (\ref{eq:imnl}) \\
$r^{sur}_{-(p,n)}$ & [fm] & 0.954 & 0.974 & (\ref{eq:imnl}) \\
$a^{sur}_{+(p,n)}$ & [fm] & 0.600 & 0.670 & (\ref{eq:imnl}) \\
$a^{sur}_{-(p,n)}$ & [fm] & 0.600 & 0.600 & (\ref{eq:imnl}) \\
$(a^{sur}_{-(p,n)})$ & [fm] & 0.500 & 0.600 & (\ref{eq:imnl}) \\
$A^{sur}_{+(p,n)}$ & [MeV] & 5.000 & 5.000 & (\ref{eq:paran2}) \\
$A^{sur}_{-(p,n)}$ & [MeV] & 1.290 & 1.204 & (\ref{eq:paran2}) \\
$B^{sur_1}_{+(p,n)}$ & [MeV] & 5.000 & 15.000 & (\ref{eq:paran2}) \\
$B^{sur_1}_{-(p,n)}$ & [MeV] & 19.431 & 23.000 & (\ref{eq:paran2}) \\
$B^{sur_2}_{+(p,n)}$ & [MeV] & 25.000 & 45.000 & (\ref{eq:paran2}) \\
$B^{sur_2}_{-(p,n)}$ & [MeV] & 60.000 & 51.000 & (\ref{eq:paran2}) \\
$C^{sur}_{+(p,n)}$ & [MeV] & 100.000 & 100.000 & (\ref{eq:paran2}) \\
$C^{sur}_{-(p,n)}$ & [MeV] & 60.000 & 51.000 & (\ref{eq:paran2}) \\
\end{tabular}
\end{ruledtabular}
\end{table}

\begin{table}[H]
\caption{Parameter values for the isoscalar part of the potential for $^{40}$Ca. The table also contains the number of the equation that defines each individual parameter.}
\label{Tbl:symmetric_40Ca}%
\begin{ruledtabular}
\begin{tabular}{ccSc} 
Parameter &  Unit & {\text{Value}} & Eq. No. \\

\hline
\multicolumn{4}{c}{Hartree-Fock} \\
\hline
$V^{HF}_{sym}$ & [MeV] & 94.820 & (\ref{eq:HFv}) \\
$(V^{HF}_{sym})$ & [MeV] & 94.020 & (\ref{eq:HFv}) \\
$\beta^{vol_1}_{sym}$ & [fm] & 1.480 & (\ref{eq:HFv}) \\
$\beta^{vol_2}_{sym}$ & [fm] & 0.728 & (\ref{eq:HFv}) \\
$x_{sym}$ &  & 0.911 & (\ref{eq:HFv}) \\
$\beta^{wb}$ & [fm] & 1.592 & (\ref{eq:wb}) \\
\hline
\multicolumn{4}{c}{Spin-orbit} \\
\hline
$a^{so}$ & [fm] & 0.755 & (\ref{eq:HFso}) \\
$\beta^{so}$ & [fm] & 0.592 & (\ref{eq:HFso}) \\
$A^{so}$ & [MeV] & -1.175 & (\ref{eq:ImSO}) \\
$B^{so}$ & [MeV] & 66.813 & (\ref{eq:ImSO}) \\
\hline
\multicolumn{4}{c}{Volume imaginary} \\
\hline
$a^{vol}_{+}$ & [fm] & 0.425 & (\ref{eq:imnl}) \\
$a^{vol}_{-}$ & [fm] & 0.253 & (\ref{eq:imnl}) \\
$B^{vol}_{+}$ & [MeV] & 29.164 & (\ref{eq:volumeS}) \\
$B^{vol}_{-}$ & [MeV] & 20.559 & (\ref{eq:volumeS}) \\
$(B^{vol}_{-})$ & [MeV] & 19.059 & (\ref{eq:volumeS}) \\
$\mathcal{E}^{vol}_{+}$ & [MeV] & 2.306 & (\ref{eq:volumeS}) \\
$\mathcal{E}^{vol}_{-}$ & [MeV] & 1.607 & (\ref{eq:volumeS}) \\
$\mathbb{E}_{+}$ & [MeV] & 31.361 & (\ref{eq:Wnmnl}) \\
$\mathbb{E}_{-}$ & [MeV] & 110.141 & (\ref{eq:Wnmnl}) \\
\hline
\multicolumn{4}{c}{Surface imaginary} \\
\hline
$a^{sur}_{+sym}$ & [fm] & 0.701 & (\ref{eq:imnl}) \\
$a^{sur}_{-sym}$ & [fm] & 0.470 & (\ref{eq:imnl}) \\
$\beta^{sur0}_{+}$ & [fm] & 1.274 & (\ref{eq:imnl}) \\
$\beta^{sur0}_{-}$ & [fm] & 1.708 & (\ref{eq:imnl}) \\
$A^{sur0}_{+}$ & [MeV] & 18.530 & (\ref{eq:paranl}) \\
$A^{sur0}_{-}$ & [MeV] & 15.023 & (\ref{eq:paranl}) \\
$B^{sur0_1}_{+}$ & [MeV] & 10.383 & (\ref{eq:paranl}) \\
$B^{sur0_1}_{-}$ & [MeV] & 10.438 & (\ref{eq:paranl}) \\
$B^{sur0_2}_{+}$ & [MeV] & 50.000 & (\ref{eq:paranl}) \\
$B^{sur0_2}_{-}$ & [MeV] & 5.941 & (\ref{eq:paranl}) \\
$C^{sur0}_{+}$ & [MeV] & 29.915 & (\ref{eq:paranl}) \\
$C^{sur0}_{-}$ & [MeV] & 14.649 & (\ref{eq:paranl}) \\
\end{tabular}
\end{ruledtabular}
\end{table}

\begin{table}[H]
\caption{Parameter values for the isoscalar part of the potential for $^{48}$Ca. The table also contains the number of the equation that defines each individual parameter.}
\label{Tbl:symmetric_48Ca}%
\begin{ruledtabular}
\begin{tabular}{ccSc} 
Parameter &  Unit & {\text{Value}} & Eq. No. \\

\hline
\multicolumn{4}{c}{Hartree-Fock} \\
\hline
$V^{HF}_{sym}$ & [MeV] & 98.226 & (\ref{eq:HFv}) \\
$(V^{HF}_{sym})$ & [MeV] & 97.726 & (\ref{eq:HFv}) \\
$\beta^{vol_1}_{sym}$ & [fm] & 1.468 & (\ref{eq:HFv}) \\
$\beta^{vol_2}_{sym}$ & [fm] & 0.708 & (\ref{eq:HFv}) \\
$x_{sym}$ &  & 0.911 & (\ref{eq:HFv}) \\
$(x_{sym})$ &  & 0.900 & (\ref{eq:HFv}) \\
$\beta^{wb}$ & [fm] & 1.592 & (\ref{eq:wb}) \\
\hline
\multicolumn{4}{c}{Spin-orbit} \\
\hline
$a^{so}$ & [fm] & 0.755 & (\ref{eq:HFso}) \\
$\beta^{so}$ & [fm] & 0.792 & (\ref{eq:HFso}) \\
$A^{so}$ & [MeV] & -1.175 & (\ref{eq:ImSO}) \\
$B^{so}$ & [MeV] & 66.813 & (\ref{eq:ImSO}) \\
\hline
\multicolumn{4}{c}{Volume imaginary} \\
\hline
$a^{vol}_{+}$ & [fm] & 0.663 & (\ref{eq:imnl}) \\
$a^{vol}_{-}$ & [fm] & 0.206 & (\ref{eq:imnl}) \\
$(a^{vol}_{-})$ & [fm] & 0.236 & (\ref{eq:imnl}) \\
$B^{vol}_{+}$ & [MeV] & 18.047 & (\ref{eq:volumeS}) \\
$B^{vol}_{-}$ & [MeV] & 20.559 & (\ref{eq:volumeS}) \\
$\mathcal{E}^{vol}_{+}$ & [MeV] & 2.306 & (\ref{eq:volumeS}) \\
$\mathcal{E}^{vol}_{-}$ & [MeV] & 1.607 & (\ref{eq:volumeS}) \\
$\mathbb{E}_{+}$ & [MeV] & 23.386 & (\ref{eq:Wnmnl}) \\
$\mathbb{E}_{-}$ & [MeV] & 97.526 & (\ref{eq:Wnmnl}) \\
$\mathbb{E}_{-}$ & [MeV] & 96.526 & (\ref{eq:Wnmnl}) \\
\hline
\multicolumn{4}{c}{Surface imaginary} \\
\hline
$a^{sur}_{+sym}$ & [fm] & 0.631 & (\ref{eq:imnl}) \\
$a^{sur}_{-sym}$ & [fm] & 0.470 & (\ref{eq:imnl}) \\
$\beta^{sur0}_{+}$ & [fm] & 1.274 & (\ref{eq:imnl}) \\
$\beta^{sur0}_{-}$ & [fm] & 1.708 & (\ref{eq:imnl}) \\
$A^{sur0}_{+}$ & [MeV] & 12.218 & (\ref{eq:paranl}) \\
$A^{sur0}_{-}$ & [MeV] & 10.023 & (\ref{eq:paranl}) \\
$B^{sur0_1}_{+}$ & [MeV] & 15.873 & (\ref{eq:paranl}) \\
$B^{sur0_1}_{-}$ & [MeV] & 7.043 & (\ref{eq:paranl}) \\
$B^{sur0_2}_{+}$ & [MeV] & 50.000 & (\ref{eq:paranl}) \\
$B^{sur0_2}_{-}$ & [MeV] & 21.073 & (\ref{eq:paranl}) \\
$C^{sur0}_{+}$ & [MeV] & 50.993 & (\ref{eq:paranl}) \\
$C^{sur0}_{-}$ & [MeV] & 16.247 & (\ref{eq:paranl}) \\
\end{tabular}
\end{ruledtabular}
\end{table}

\bibliographystyle{apsrev4-1}
\bibliography{Numerical}

\end{document}